\documentstyle[aps,notebook]{revtex}
\def\CellGroup{\bgroup}
\def\endCellGroup{\egroup}

\skewchar\fivmi='177
\skewchar\sixmi='177
\skewchar\sevmi='177
\skewchar\egtmi='177
\skewchar\ninmi='177
\skewchar\tenmi='177
\skewchar\elvmi='177
\skewchar\twlmi='177
\skewchar\frtnmi='177
\skewchar\svtnmi='177
\skewchar\twtymi='177
\def\@magscale#1{ scaled \magstep #1}
\skewchar\fivsy='60
\skewchar\sixsy='60
\skewchar\sevsy='60
\skewchar\egtsy='60
\skewchar\ninsy='60
\skewchar\tensy='60
\skewchar\elvsy='60
\skewchar\twlsy='60
\skewchar\frtnsy='60
\skewchar\svtnsy='60
\skewchar\twtysy='60

\catcode`@=11
\def\un#1{\relax\ifmmode\@@underline#1\else
        $\@@underline{\hbox{#1}}$\relax\fi}
\catcode`@=12

\def\a{\alpha}
\def\b{\beta}
\def\c{\chi}
\def\d{\delta}
\def\e{\epsilon}

\def\g{\gamma}
\def\h{\eta}

\def\j{\psi}

\def\s{\sigma}

\def\dslash{\not{\hbox{\kern-2pt $\partial$}}}
\def\Dslash{\not{\hbox{\kern-4pt $D$}}}
\def\pslash{\not{\hbox{\kern-2.3pt $p$}}}
 \newtoks\slashfraction
 \slashfraction={.13}
 \def\slash#1{\setbox0\hbox{$ #1 $}
 \setbox0\hbox to \the\slashfraction\wd0{\hss \box0}/\box0 }
 
\font\ro=cmsy10                          % font with rope
\def\kcr{{\hbox{\ro \char'170}}}                % right-handed rope
\def\ktl{{\hbox{\ro \char'170}}}        % top end for left-handed rope
\def\ktr{{\hbox{\ro \char'170}}}        % " right
\def\kbl{{\hbox{\ro \char'170}}}        % " bottom left
\def\kbr{{\hbox{\ro \char'170}}}        % " right
\def\plpl{\raise-2pt\hbox{$\raise3pt\hbox{$_+$}\hskip-6.67pt\raise0.0pt
\hbox{$^+$}\hskip 0.01pt$}}
\def\mimi{\raise-2pt\hbox{$\raise3pt\hbox{$_-$}\hskip-6.67pt\raise0.0pt
\hbox{$^-$}\hskip 0.01pt$}} 

\def\bo{{\raise.15ex\hbox{\large$\Box$}}}               % D'Alembertian
\def\TH{{\raise.2ex\hbox{$\displaystyle \bigodot$}\mskip-4.7mu \llap H \;}}
\def\face{{\raise.2ex\hbox{$\displaystyle \bigodot$}\mskip-2.2mu \llap {$\ddot
        \smile$}}}                                      % happy face
\def\sp#1{{}^{#1}}                              % superscript (unaligned)
\def\leftrightarrowfill{$\mathsurround=0pt \mathord\leftarrow \mkern-6mu
        \cleaders\hbox{$\mkern-2mu \mathord- \mkern-2mu$}\hfill
        \mkern-6mu \mathord\rightarrow$}
\def\dvec#1{\vbox{\ialign{##\crcr
        \leftrightarrowfill\crcr\noalign{\kern-1pt\nointerlineskip}
        $\hfil\displaystyle{#1}\hfil$\crcr}}}           % <--> accent
\def\frac#1#2{{\textstyle{#1\over\vphantom2\smash{\raise.20ex
        \hbox{$\scriptstyle{#2}$}}}}}                   % fraction
\def\sfrac#1#2{{\vphantom1\smash{\lower.5ex\hbox{\small$#1$}}\over
        \vphantom1\smash{\raise.4ex\hbox{\small$#2$}}}} % alternate fraction
\def\bfrac#1#2{{\vphantom1\smash{\lower.5ex\hbox{$#1$}}\over
        \vphantom1\smash{\raise.3ex\hbox{$#2$}}}}       % "
\def\afrac#1#2{{\vphantom1\smash{\lower.5ex\hbox{$#1$}}\over#2}}    % "
\newskip\humongous \humongous=0pt plus 1000pt minus 1000pt

\newif\ifdtup

\def\ref#1{$\sp{#1)}$}

\parskip=\medskipamount                 % space between paragraphs (LaTeX)
\thicklines                         % thick straight lines for pictures (LaTeX)

\def\oldheadpic{                                % old UM heading
        \setlength{\unitlength}{.4mm}
        \thinlines
        \par
        \begin{picture}(349,16)
        \put(325,16){\line(1,0){4}}
        \put(330,16){\line(1,0){4}}
        \put(340,16){\line(1,0){4}}
        \put(335,0){\line(1,0){4}}
        \put(340,0){\line(1,0){4}}
        \put(345,0){\line(1,0){4}}
        \put(329,0){\line(0,1){16}}
        \put(330,0){\line(0,1){16}}
        \put(339,0){\line(0,1){16}}
        \put(340,0){\line(0,1){16}}
        \put(344,0){\line(0,1){16}}
        \put(345,0){\line(0,1){16}}
        \put(329,16){\oval(8,32)[bl]}
        \put(330,16){\oval(8,32)[br]}
        \put(339,0){\oval(8,32)[tl]}
        \put(345,0){\oval(8,32)[tr]}
        \end{picture}
        \par
        \thicklines
        \vskip.2in}
\def\oldtitle#1#2#3#4{\oldheadpic\begin{center}\vglue.5in{\large\bf #1}\\[.6in]
        {#2}\\[.1in] {\it Department of Physics and Astronomy}\\
        {\it University of Maryland, College Park, MD 20742}\\[.6in]
        Physics Publication \#{#3}\\ {#4}\\[1.5in] {\bf ABSTRACT}\\[.1in]
        \end{center} \begin{quotation}}                 % old title stuff
\def\oldTitle#1#2#3#4#5#6#7{\oldheadpic\begin{center} \vglue .4in
        {\large\bf #1}\\[.4in]
        {#2}\\[.1in] {\it Department of Physics and Astronomy}\\
        {\it University of Maryland, College Park, MD 20742}\\[.1in]
        {#3}\\[.1in] {\it {#4}}\\ {\it {#5}}\\[.4in]
        Physics Publication \#{#6}\\ {#7}\\[.5in] {\bf ABSTRACT}\\[.1in]
        \end{center} \begin{quotation}}                 % " for 2 authors
\def\border{                                            % border
        \setlength{\unitlength}{1mm}
        \newcount\xco
        \newcount\yco
        \xco=-21
        \yco=12
        \begin{picture}(140,0)
        \put(\xco,\yco){$\ktl$}
        \advance\yco by-1
        {\loop
        \put(\xco,\yco){$\kcr$}
        \advance\yco by-2
        \ifnum\yco>-240
        \repeat
        \put(\xco,\yco){$\kbl$}}
        \xco=158
        \yco=12
        \put(\xco,\yco){$\ktr$}
        \advance\yco by-1
        {\loop
        \put(\xco,\yco){$\kcr$}
        \advance\yco by-2
        \ifnum\yco>-240
        \repeat
        \put(\xco,\yco){$\kbr$}}
        \put(-20,13){\tiny University of Maryland Elementary Particle
Physics University of Maryland Elementary Particle Physics University of
Maryland Elementary Particle Physics}
        \put(-20,-241.5){\tiny University of Maryland Elementary
Particle Physics University of Maryland Elementary Particle Physics
University of Maryland Elementary Particle Physics}
        \end{picture}
        \par\vskip-8mm}
\def\bordero{                                           % alternate border
        \setlength{\unitlength}{1mm}
        \newcount\xco
        \newcount\yco
        \xco=-31
        \yco=12
        \begin{picture}(140,0)
        \put(\xco,\yco){$\ktl$}
        \advance\yco by-1
        {\loop
        \put(\xco,\yco){$\kclr$}
        \advance\yco by-2
        \ifnum\yco>-240
        \repeat
        \put(\xco,\yco){$\kbl$}}
        \xco=151
        \yco=12
        \put(\xco,\yco){$\ktr$}
        \advance\yco by-1
        {\loop
        \put(\xco,\yco){$\kcr$}
        \advance\yco by-2
        \ifnum\yco>-240
        \repeat
        \put(\xco,\yco){$\kbr$}}
        \put(-20,12){\ooo bacdefghidfghghdhededbihdgdfdfhhdheidhdhebaaahjhhdahba

hgdedge
   hgfdiehhgdigicba}
        \put(-20,-241.5){\ooo ababaighefdbfghgeahgdfgafagihdidihiidhiagfedhadbfd

ecdcdfa
   gdcbhaddhbgfchbgfdacfediacbabab}
        \end{picture}
        \par\vskip-8mm}
\def\headpic{                                           % UM heading
        \indent
        \setlength{\unitlength}{.4mm}
        \thinlines
        \par
        \begin{picture}(29,16)
        \put(165,16){\line(1,0){4}}
        \put(170,16){\line(1,0){4}}
        \put(180,16){\line(1,0){4}}
        \put(175,0){\line(1,0){4}}
        \put(180,0){\line(1,0){4}}
        \put(185,0){\line(1,0){4}}
        \put(169,0){\line(0,1){16}}
        \put(170,0){\line(0,1){16}}
        \put(179,0){\line(0,1){16}}
        \put(180,0){\line(0,1){16}}
        \put(184,0){\line(0,1){16}}
        \put(185,0){\line(0,1){16}}
        \put(169,16){\oval(8,32)[bl]}
        \put(170,16){\oval(8,32)[br]}
        \put(179,0){\oval(8,32)[tl]}
        \put(185,0){\oval(8,32)[tr]}
        \end{picture}
        \par\vskip-6.5mm
        \thicklines}
\def\title#1#2#3#4{\border\headpic {\hbox to\hsize{#4 \hfill UMDEPP #3}}\par
        \begin{center} \vglue .5in {\large\bf #1}\\[.6in]
        {#2}\\[.1in] {\it Department of Physics and Astronomy}\\
        {\it University of Maryland, College Park, MD 20742}\\[1.5in]
        {\bf ABSTRACT}\\[.1in] \end{center} \begin{quotation}}  % title stuff
\def\Title#1#2#3#4#5#6#7{\border\headpic
        {\hbox to\hsize{#7 \hfill UMDEPP #6}}\par
        \begin{center} \vglue .4in {\large\bf #1}\\[.4in]
        {#2}\\[.1in] {\it Department of Physics and Astronomy}\\
        {\it University of Maryland, College Park, MD 20742}\\[.1in]
        {#3}\\[.1in] {\it {#4}}\\ {\it {#5}}\\[.5in] {\bf ABSTRACT}\\[.1in]
        \end{center} \begin{quotation}}                 % " for 2 authors
\def\endtitle{\end{quotation}\newpage}                  % end title page

\def\sect#1{\bigskip\medskip \goodbreak \noindent{\bf {#1}} \nobreak \medskip}
\def\refs{\sect{REFERENCES} \footnotesize \frenchspacing \parskip=0pt}

\begin{document}
\par
\setlength{\oddsidemargin}{0.3in}
\setlength{\evensidemargin}{-0.3in}
\begin{center}
\vglue .15in
{\large\bf Irreducible Decomposition of Products of 10D Chiral Sigma Matrices
\footnote {Supported 
in part by National  Science Foundation Grant PHY-98-02551.}  }
\\[.5in]
S. James Gates, Jr.\footnote{gatess@wam.umd.edu} and
B. Radak\footnote{radakbra@fsg.prusec.com}
\\[0.06in]
{\it Department of Physics, 
University of Maryland\\ 
College Park, MD 20742-4111  USA}\\[.1in]
and \\ [.1in] 
V.G.J. Rodgers\footnote{vincent-rodgers@uiowa.edu }
\\[0.06in]
{\it  Department of Physics and Astronomy, 
University of Iowa\\ 
Iowa City, Iowa~~52242--1479 USA}\\[.9in]

{\bf ABSTRACT}\\[.01in]
\end{center}
\begin{quotation}
We review the enveloping algebra of the 10 dimensional chiral sigma matrices.
To facilitate the computation of the product of several
chiral sigma matrices we have developed a symbolic program.  Using this
program one can reduce the multiplication of the sigma matrices down to
linear combinations of irreducilbe elements.  We are able to quickly
derive several identities that are not restricted to traces. 
A copy of the program written in the Mathematica language is provided
for the community. 
\endtitle

\section{Spinors and Sigma matrices in 10 D}
\label{sec:1}
Several problems in 10 D Physics require the ability to reduce 
products of the 10 D  chiral sigma matrices into a 
linear combination of the irreducible matrices [1-4].   In general
such calculations can quickly become intractable.  For some problems,
many approximation 
schemes are applied such as linearization and perturbation theory.
However when one is discussing properties of algebras, such as the 
closing of the Jacobi identity or commutation relations, exact results 
are mandatory [5].  Computations such as computing the equations of
motion in some supergravity theories, could literally take several 
weeks.  These types of calculations could be reduced to a few hours 
work using today's powerful symbolic 
manipulators, such as Mathematica and Maple,  in conjuction with 
high speed computers. 
In this note we provide the community with an example of such a
program along with several identities that we have derived using
this tool.  We begin by outlining the formal construction of the
enveloping algebra of the 10 dimensional chiral sigma matrices.

It is well known that in D-dimensional space-time Dirac spinor has 
$2^{D\over 2}$ 
complex components if D is even and in that case all the representations 
of the Dirac algebra are equivalent, while in odd dimensional space-time the
number of components is  $2^{{D-1}\over 2}$ with two inequivalent
representations. This result is valid for any signature of the flat
space-time metric.  However, this kind of spinor is not, for any D,
the smallest
irreducible representation.  When D is even we can always impose the Weyl
condition which separates the spinor into its left and right handed
components which transform independently under the Lorentz group. Each of
these Weyl spinors thus has only half degrees of freedom of the original
Dirac spinor. One more condition can be imposed to reduce further the number
of independent components. This is so called Majorana condition which relates
the charge-conjugate Dirac spinor to the original one. If after the charge
conjugation the spinor is not changed, the number of independent components
is reduced again by a factor of two. In that case the left and right Weyl
spinors are related. The resulting spinor is necessarily real. This type of
reduction can be done in 2 (mod 8) dimensional space-time. It then follows
that for $D=10~$ we can impose both conditions and have for the basic spinor
a 16-dimensional real object.  Consequently, instead of using the Dirac gamma
matrices we use the Pauli matrices, of dimension $16\times 16~$ with real
elements. We can then use small Greek index to denote 16 component
left-handed Majorana-Weyl spinor and dotted index for a right-handed one:
\begin{equation}(\j^{\a})^{*}~=~(\j^{\a}),~~~~~(\c^{\dot{\a}})^{*}~=~(\c^{\dot{\a}}).
\label{eqno(1)}
\end{equation}
The indices can be raised and lowered by the use of the charge conjugation 
matrix which has mixed indices: $C_{\a\dot\b}~$ and $C^{\a\dot\b}$,
\begin{eqnarray}
&&\j_{\dot\b}~=~\j^{\a}C_{\a\dot\b},~~~~~~~~\c_{\a}~=~\c^{\dot\b}C_{\a\dot\b},
\\
&&C_{\a\dot\g}C^{\a\dot\b}~=~\d_{\dot\g}^{\dot\b},~~~~~
C^{\a\dot\b}C_{\g\dot\b}~=~\d_{\g}^{\a}. \label{eqno(2)}
\end{eqnarray}
As we can see, raising and lowering the indices also changes the handedness 
of a spinor. The Dirac algebra (of the Pauli matrices) is defined through 
the usual anticommutation relation:
\begin{equation}
(\s_{\un a})_{\a\b}(\s_{\un b})^{\b\g}~+~(\s_{\un b})_{\a\b}(\s_{\un a})^{\b
\g}~=~-2\h_{\un a \un b}\d_{\a}^{\g}~~.\label{eqno(3)}
\end{equation}
The sigma matrices can be regarded as bispinors. There are three types of them:
purely left-handed, purely right-handed and mixed ones. The purely left-handed 
ones are
\begin{equation}
(\s^{\un a})_{\a\b}, ~~~(\s^{\un a \un b \un c})_{\a\b},~~~~(\s^{
{\un a}_{1} {\un a}_{2} {\un a}_{3} {\un a}_{4} {\un a}_{5}})_{\a\b}
,\label{4}
\end{equation}
where lower case Latin indices are vector indices denoting space-time 
directions. Purely right-handed bispinors have only dotted indices, but due to 
the existence of the charge conjugation matrix we have
\begin{equation}
(\s^{\un a})^{\a\b}~=~C^{\a\dot\a}C^{\b\dot\b}(\s^{\un a})_{\dot\a
\dot\b}.\label{5}
\end{equation}
The mixed bispinors are 
\begin{equation}
C_{\a\dot\b},~~~~(\s^{\un a \un b})_{\a\dot\b},~~~~(\s^{\un a \un b
\un c \un d})_{\a\dot\b}.\label{eqno(6)}
\end{equation}
They are, of course, related to  
\begin{equation}
\d_{\a}^{~\b},~~~(\s^{\un a \un b})_{\a}^{~\b},~~~~(\s^{\un a \un b 
\un c \un d})_{\a}^{~\b}.\label{eqno(7)}
\end{equation}
 
The sigma matrices with more vector indices are defined through the following 
multiplication table:
\begin{eqnarray}
(\s_{\un a})_{\a\b}(\s_{\un b})^{\b\g}~&=&~-\h_{\un a \un b}\d_{\a}^{~\g}
~-~(\s_{\un a \un b})_{\a}^{~\g},
\nonumber   \\
(\s_{\un a})_{\a\b}(\s_{\un b \un c})^{~\b}_{\g}~&=&~-\h_{\un a [\un b}
(\s_{\un c]})_{\a\g}~-~(\s_{\un a \un b \un c})_{\a\g},\nonumber \\
(\s_{\un a})_{\a\b}(\s_{\un b \un c \un d})^{\b\g}~&=&~-{1\over {2!}}
\h_{\un a[\un b}(\s_{\un c \un d]})_{\a}^{~\g}~-~(\s_{\un a \un b 
\un c \un d})_{\a}^{~\g},\label{eqno(8)}\\
(\s_{\un a})_{\a\b }(\s_{\un b \un c \un d \un e})^{~\b}_{\g}~&=&~
{1\over {3!}}\h_{\un a[\un b}
(\s_{\un c \un d \un e]})_{\a \g}~+~(\s_{\un a \un b \un c \un d \un e})_{\a
\g},\nonumber\\
(\s_{\un a})_{\a\b}(\s_{\un b \un c \un d \un e \un f})^{\b\g}~&=&~-{1\over
{4!}}\h_{\un a[\un b}(\s_{\un c \un d \un e \un f]})_{\a}^{~\g}~-~
{1\over {4!}}\e_{\un a \un b \un c \un d \un e \un
  f[4]}(\s^{[4]})_{\a}^{~\g}.
\nonumber
\end{eqnarray}

\begin{eqnarray}
(\s_{\un a})_{\a\b}(\s_{\un b})^{\b\g}~&=&~-\h_{\un a \un b}\d_{\a}^{~\g}~
-~(\s_{\un a \un b})_{\a}^{~\g},
\nonumber\\
(\s_{\un b \un c})^{~\a}_{\b}(\s_{\un a})^{\b\g}~&=&~+\h_{\un a [ \un b}
(\s_{\un c]})^{\a\g}~-~(\s_{\un a \un b \un c})^{\a\g},\nonumber\\
(\s_{\un b \un c \un d})_{\a\b}(\s_{\un a})^{\b\g}~&=&~-{1\over {2!}}
\h_{\un a [ \un b}(\s_{\un c \un d ]})_{\a}^{~\g}~+~(\s_{\un a \un b \un c
\un d})_{\a}^{~\g},\label{eqno(9)}\\
(\s_{\un b \un c \un d \un e})^{~\a}_{\b}(\s_{\un a})^{\b\g}~&=&~-{1
\over {3!}}\h_{\un a [ \un b}(\s_{\un c \un d \un e]})^{\a\g}~+~
(\s_{\un a \un b \un c \un d \un e})^{\a\g},\nonumber\\
(\s_{\un b \un c \un d \un e \un f})_{\a\b}(\s_{\un a})^{\b\g}~&=&~-{1
\over {4!}}\h_{\un a [ \un b}(\s_{\un c \un d \un e \un f]})_{
\a}^{~\g}~+~{1\over {4!}}\e_{\un a \un b \un c \un d \un e \un f [4]}
(\s^{[4]})_{\a}^{~\g}.
\nonumber
\end{eqnarray}

All of the sigma matrices are totally antisymmetric in their vector indices.
The sigma matrices with one or five vector indices are symmetric with respect 
to spinor indices, while the matrix with three vector indices is antisymmetric 
in spinor indices. The symmetrization and antisymmetrization is denoted in 
the following way:
\begin{eqnarray}
A_{(\a}B_{\b)}~&=&~A_{\a}B_{\b}~+~A_{\b}B_{\a},\nonumber\\
A_{[\a}B_{\b]}~&=&~A_{\a}B_{\b}~-~A_{\b}B_{\a}.\label{eqno(10)}
\end{eqnarray}
The sigma matrices with five vector indices satisfy the identities:
\begin{eqnarray}
&&(\s_{[5]})_{\a\b}~=~~{1\over {5!}}\e_{[5]\bar{[5]}}(\s^{\bar{[5]}})_{\a\b},
\nonumber\\
&&(\s_{[5]})^{\a\b}~=~-{1\over {5!}}\e_{[5]\bar{[5]}}(\s^{\bar{[5]}})^{\a\b}.
\label{eqno(11)}
\end{eqnarray}
It is also useful to define matrices with more than five indices
 
\begin{eqnarray}
(\s_{[6]})_{\a}^{~\b}~&=&~~{1\over {4!}}\e_{[6]\bar{[4]}}(\s^{\bar{[4]}})_{\a}
^{~\b},\nonumber\\
(\s_{[7]})_{\a\b}~&=&~-{1\over {3!}}\e_{[7]\bar{[3]}}(\s^{\bar{[3]}})_{\a\b},
\nonumber\\
(\s_{[7]})^{\a\b}~&=&~-{1\over {3!}}\e_{[7]\bar{[3]}}(\s^{\bar{[3]}})^{\a\b},
\nonumber\\
(\s_{[8]})_{\a}^{~\b}~&=&~-{1\over {2!}}\e_{[8]\bar{[2]}}(\s^{\bar{[2]}})_{\a}
^{~\b},\label{eqno(12)}\\
(\s_{[9]})_{\a\b}~&=&~~\e_{[9]\un a}(\s^{\un a})_{\a\b},\nonumber\\
(\s_{[9]})^{\a\b}~&=&~~\e_{[9] \un a}(\s^{\un
  a})^{\a\b}. \nonumber
\end{eqnarray}

The trace identities are:
\begin{eqnarray}
&&(\s_{\un a})_{\a\b}(\s^{\un c})^{\a\b}~=~
-16~\d_{\un a}^{\un c}~~, \nonumber\\
&&(\s_{\un a \un b})_{\a}^{~\b}(\s^{\un c \un d})_{\b}^{~\a}~=~
-16~\d_{[\un a}^{\un c} \d_{\un b]}^{\un d}~~,\nonumber \\
&&(\s_{\un a \un b \un c})_{\a\b}(\s^{\un d \un e \un f})^{\a\b}~=~
-16~\d_{[\un a}^{\un d} \d_{\un b}^{\un e} \d_{\un c]}^{\un f} ~~,\label{eqno(13)} \\
&&(\s_{\un a \un b \un c \un d})_{\a}^{~\b}(\s^{\un e \un f \un g \un h})_{
\b}^{~\a}~=~16~\d_{[\un a}^{\un e} \d_{\un b}^{\un f} \d_{\un c}^{\un g}
\d_{\un d]}^{\un h} ~~ ,\nonumber \\
&&(\s_{\un a \un b \un c \un d \un e})_{\a\b}(\s^{\un f \un g \un h \un i
\un j})^{\a\b}~=~-16~[\d_{[\un a}^{\un f} \d_{\un b}^{\un g} \d_{\un c}^{\un h}
\d_{\un d}^{\un i} \d_{\un e]}^{\un j}~+~\e_{\un a \un b \un c \un d
\un e}^{~~~~~\un f \un g \un h \un i \un j}].
\nonumber
\end{eqnarray}
 
In terms of ordinary matrices, our notation can be regarded as follows.
We can introduce a bar-notation to distinguish the altitude of the
indices on the bi-spinors in Eq(5). In otherwords we define the matrices
\begin{equation}
 \s_{\un a} \equiv (\s_{\un a})_{\a \b} ~~~~,~~~~
{\bar \s}_{\un a} \equiv (\s_{\un a})^{\a \b} ~~~, \label{eqno(14)}
\end{equation}
and similarly for the third and fifth rank space-time tensors.We further use
a Northwest-Southeast convention with respect to the contraction of hidden
spinor indices. This is simplest to understand via the following examples
\begin{equation}
\s_{\un a} {\bar \s}_{\un b} \equiv - (\s_{\un a})_{\a \b} (\s_{
\un b})^{\b \g} ~~~~, ~~~~
 {\bar \s}_{\un b} \s_{\un a} \equiv (\s_{ \un b})^{\g \b} (\s_{
\un a})_{\b \a}  ~~~~.  
\end{equation}
So when the contraction is up-to-down there is a plus sign when writing out 
the explicit indices, but when the contraction is down-to-up there is a
minus sign.

The sigma matrices with more vector indices follow from
the multiplication table:
\begin{eqnarray}
\s_{\un a} {\bar \s}_{\un b} ~&&=~\h_{\un a \un b} ~+~ \s_{\un a \un b} ~~~,
\nonumber   \\
{\bar \s}_{\un a} \s_{\un b} ~&&=~- \h_{\un a \un b} ~+~ \s_{\un a \un b} ~~~,
\nonumber   \\
\s_{\un a} \s_{\un b \un c} ~&&=~ \h_{\un a [\un b}
\s_{\un c]} ~+~ \s_{\un a \un b \un c} ~~~,\nonumber\\
\s_{\un b \un c} \s_{\un a}  ~&&=~ - \h_{\un a [\un b}
\s_{\un c]} ~+~ \s_{\un a \un b \un c} ~~~,\nonumber\\
\s_{\un a} {\bar \s}_{\un b \un c \un d} ~&&=~ {1\over {2!}}
\h_{\un a[\un b} \s_{\un c \un d]} ~+~ \s_{\un a \un b 
\un c \un d} ~~~,\nonumber\\
{\bar \s}_{\un b \un c \un d} \s_{\un a} ~&&=~ {1\over {2!}}
\h_{\un a[\un b} \s_{\un c \un d]} ~+~ \s_{\un a \un b 
\un c \un d} ~~~,\label{eqno(16)}\\
\s_{\un a} \s_{\un b \un c \un d \un e} ~&&=~ -
{1\over {3!}}\h_{\un a[\un b} \s_{\un c \un d \un e]} ~-~ \s_{\un a \un b
\un c \un d \un e} ~~~,\nonumber \\
\s_{\un b \un c \un d \un e} \s_{\un a}  ~&&=~ -
{1\over {3!}}\h_{\un a[\un b} \s_{\un c \un d \un e]} ~+~ \s_{\un a \un b
\un c \un d \un e} ~~~,\nonumber\\
\s_{\un a} {\bar \s}_{\un b \un c \un d \un e \un f} ~&&=~ {1\over
{4!}}\h_{\un a[\un b} \s_{\un c \un d \un e \un f]} ~+~
{1\over {4!}}\e_{\un a \un b \un c \un d \un e \un f[4]}\s^{[4]}~~~
,\nonumber\\
{\bar \s}_{\un b \un c \un d \un e \un f} \s_{\un a} ~&&=~ - {1\over
{4!}}\h_{\un a[\un b} \s_{\un c \un d \un e \un f]} ~-~
{1\over {4!}}\e_{\un a \un b \un c \un d \un e \un f[4]}\s^{[4]}~~~.
\nonumber
\end{eqnarray}

\begin{eqnarray}
{\bar \s}_{\un a} \s_{\un b \un c} ~&&=~ \h_{\un a [\un b}
{\bar \s}_{\un c]} ~+~ {\bar \s}_{\un a \un b \un c} ~~~,\nonumber\\
\s_{\un b \un c} {\bar \s}_{\un a}  ~&&=~ - \h_{\un a [\un b}
{\bar \s}_{\un c]} ~+~ {\bar \s}_{\un a \un b \un c} ~~~,\nonumber\\
{\bar \s}_{\un a} \s_{\un b \un c \un d} ~&&=~ {1\over {2!}}
\h_{\un a[\un b} \s_{\un c \un d]} ~-~ \s_{\un a \un b 
\un c \un d} ~~~,\nonumber\\
\s_{\un b \un c \un d} {\bar \s}_{\un a} ~&&=~ {1\over {2!}} \h_{\un a[\un b}
 \s_{\un c \un d]} ~-~ \s_{\un a \un b \un c \un d} ~~~,\nonumber\\
{\bar \s}_{\un a} \s_{\un b \un c \un d \un e} ~&&=~ 
{1\over {3!}}\h_{\un a[\un b} {\bar \s}_{\un c \un d \un e]} ~+~ 
{\bar \s}_{\un a \un b\un c \un d \un e} ~~~,\label{eqno(17)}\\
\s_{\un b \un c \un d \un e} {\bar \s}_{\un a}  ~&&=~ 
{1\over {3!}}\h_{\un a[\un b} {\bar \s}_{\un c \un d \un e]} ~-~ 
{\bar \s}_{\un a \un b \un c \un d \un e} ~~~,\nonumber\\
{\bar \s}_{\un a} \s_{\un b \un c \un d \un e \un f} ~&&=~ - {1\over
{4!}}\h_{\un a[\un b} \s_{\un c \un d \un e \un f]} ~+~
{1\over {4!}}\e_{\un a \un b \un c \un d \un e \un f[4]}\s^{[4]}~~~
,\nonumber\\
\s_{\un b \un c \un d \un e \un f} {\bar \s}_{\un a} ~&&=~  {1\over
{4!}}\h_{\un a[\un b} \s_{\un c \un d \un e \un f]} ~-~
{1\over {4!}}\e_{\un a \un b \un c \un d \un e \un f[4]}\s^{[4]}~~~.\nonumber
\end{eqnarray}

This establishes the enveloping algebra of the 10 D chiral matrices.
It is easy to see that one can use these rules to reduce any product of
such matrices down to a linear combination of irreducible elements.  
However the presence of the two copies of five forms down to one forms
renders many of the calculatons intractable.  In what follows we will
introduce a program using the Mathematica language that can facilitate
such computations.  We will first give an example worksheet that shows
how to use the program then present results that have exploited the
routine.  In the last section we present the program verbatim.  We
will call the program SigmaVector10D.m.  

\section{Instructions for Using  SigmaVector10D.m}
\label{sec:2}

%% 
%% This is a LaTeX document generated by automatic conversion
%% from a 
%% Mathematica notebook using Mathematica 3.0.
%% 
%% This document uses special macros defined in the style file
%% notebook.sty.
%% To run the document, you must put this style file in a
%% directory on
%% the path searched by LaTeX.  The style file is in the
%% IncludeFiles
%% subdirectory of the SystemFiles directory in your
%% Mathematica distribution.
%% 

\def\CellGroup{\bgroup}
\def\endCellGroup{\egroup}

%%\documentstyle[notebook]{article}

%%\begin{document}

Here is an example run of the Mathematica package
SigmaVector10D.m. 
First one must copy the program over to a Mathematica notebook
and save the file as {\bf SigmaVector10D.m}.  Then open another 
notebook and bring in this file by typing,
``{\bf $<<$\ SigmaVector.m}''.  Below is a transcript of a typical
run.
\begin{CellGroup}
\mathin
<<SigmaVector10D.m
\endmathin
\mathout
\endmathout
This is SigmaVector program. It reduces the product of
sigma matrices. Spinor indices of the chiral sigma 
matrices analyzed here are supressed. Sigma matrices
with spinor indices down are denoted by S, as well as
those  matrices which have one spinor index down and
other up. If both spinor indices are up, the letter SB
is used. The program handles all contraction
possibilities of spinor indices, except when two sigma
matrices, both with even number of vector indices are
multiplied in such a way that the lower index of the
first sigma matrix is summed with the upper index of
the second matrix. In this case the order of sigma
matrices should be reversed. Here is an example:
$(\s^{e f})_{\a}^{\b}*(\s_{a b c})^{\a\g}*(\s_{e f})_{\g}^{\d}$
should  be  represented  in  the  symbolic  form  as:
$ {\bf ExpandAll[ExpandAll[S[e,f]**SB[a,b,c]]**S[e,f]//.Rule1],}$
where  Rule1  is  a  rule  which  transforms  sigma  matrices
with  6  or  more  vector  indices  in  appropriate  sigma
matrices  with  4  or  less  vector  indices  and  contracts
corresponding  epsilon  symbols.
%\endmathout
\end{CellGroup}
\mathin
Here\ is\ the\ first\ example:
\endmathin
\begin{CellGroup}
\dispSFinmath{\Mfunction{Expand}[S[a,b]**S[a,b,c]]}
\dispSFoutmath{-2\ S[c]+3\ \Mvariable{Ten}\ S[c]-{{\Mvariable{Ten}}^2}\ S[c]}

\end{CellGroup}
\mathin
The\ output\ tells\ you\ something\ about\ the\ Kroneker\ 

delta\ symbols.\ In\ most\ cases\ you\ are\ not\ interested\ in\ 

that,\ so\ you\ type:
\endmathin
\begin{CellGroup}
\dispSFinmath{\Mvariable{Ten}=10}
\dispSFoutmath{10}

\end{CellGroup}
\mathin
The\ previous\ example\ now\ becomes:
\endmathin
\begin{CellGroup}
\dispSFinmath{\Mfunction{Expand}[S[a,b]**S[a,b,c]]}
\dispSFoutmath{-72\ S[c]}

\end{CellGroup}
\begin{CellGroup}
\dispSFinmath{\MathBegin{MathArray}{l}
\Mvariable{ExpandAll}[  \\
\noalign{\vspace{0.555556ex}}
\hspace{1.em} \Mfunction{ExpandAll}[S[e,f]**S[a]]**S[e,f]]\\
\MathEnd{MathArray}}
\dispSFoutmath{54\ S[a]}

\end{CellGroup}
\mathin

\endmathin
\mathin
Note\ that\ we\ have\ grouped\ the\ products\ of\ sigma\ matrices.\ 

This\ reduces\ the\ run\ time\ by\ a\ factor\ of\ three!\ When\ you\ 

have\ more\ than\ 10\ vector\ indices,\ this\ is\ very\ important.\

In\ this\ simple\ example\ we\ could\ have\ used\ the\ following\ 

command:
\endmathin
\begin{CellGroup}
\dispSFinmath{S[e,f]**S[a]**S[e,f]}
\dispSFoutmath{54\ S[a]}

\end{CellGroup}
\mathin
Here\ comes\ a\ nontrivial\ example:
\endmathin
\begin{CellGroup}
\dispSFinmath{\MathBegin{MathArray}{l}
\Mvariable{ExpandAll}[  \\
\noalign{\vspace{0.555556ex}}
\hspace{1.em} \Mfunction{ExpandAll}[\Muserfunction{SB}[e,f,g]**S[a,b,c]]**  \\
\noalign{\vspace{0.555556ex}}
\hspace{2.em} \Muserfunction{SB}[e,f,g]]\\
\MathEnd{MathArray}}
\dispSFoutmath{\MathBegin{MathArray}{l}
258\ \Muserfunction{SB}[a,b,c]+\frac{1}{24}\ \Muserfunction{Epsilon}[a,b,c,e,  \\
\noalign{\vspace{0.847222ex}}
\hspace{3.em} f,g,\$1[1],\$2[1],\$3[1],\$4[1]]\   \\
\noalign{\vspace{0.555556ex}}
\hspace{2.em} \Muserfunction{SB}[e,f,g,\$1[1],\$2[1],\$3[1],\$4[1]]\\
\MathEnd{MathArray}}

\end{CellGroup}
\mathin
We\ have\ obtained\ a\ sigma\ matrix\ with\ 7\ vector\ indices!

The\ introduction\ of\ such\ matrices\ is\ important\ -\ the\ run\ 

time\ is\ much\ shorter.\ We\ now\ want\ to\ transform\ this\ 

7-index\ matrix\ into\ a\ product\ of\ epsilon\ symbol\ and\ a\ 

matrix\ with\ 3\ indices\ by\ the\ use\ of\ "Rule1".\ Two\ epsilon\ 

symbols\ will\ be\ automatically\ contracted.
\endmathin
\begin{CellGroup}
\dispSFinmath{\%//.\Mvariable{Rule1}}
\dispSFoutmath{\MathBegin{MathArray}{l}
258\ \Muserfunction{SB}[a,b,c]-  \\
\noalign{\vspace{0.888889ex}}
\hspace{1.em} \frac{1}{144}\ (-5040\ \Muserfunction{Id}[a,\$3[6]]\ \Muserfunction{Id}[b,\$2[6]]\   \\
\noalign{\vspace{0.847222ex}}
\hspace{5.em} \Muserfunction{Id}[c,\$1[6]]+5040\ \Muserfunction{Id}[a,\$2[6]]\   \\
\noalign{\vspace{0.555556ex}}
\hspace{5.em} \Muserfunction{Id}[b,\$3[6]]\ \Muserfunction{Id}[c,\$1[6]]+  \\
\noalign{\vspace{0.555556ex}}
\hspace{4.em} 5040\ \Muserfunction{Id}[a,\$3[6]]\   \\
\noalign{\vspace{0.555556ex}}
\hspace{5.em} \Muserfunction{Id}[b,\$1[6]]\ \Muserfunction{Id}[c,\$2[6]]-  \\
\noalign{\vspace{0.555556ex}}
\hspace{4.em} 5040\ \Muserfunction{Id}[a,\$1[6]]\ \Muserfunction{Id}[b,\$3[6]]\   \\
\noalign{\vspace{0.555556ex}}
\hspace{5.em} \Muserfunction{Id}[c,\$2[6]]-5040\ \Muserfunction{Id}[a,\$2[6]]\   \\
\noalign{\vspace{0.555556ex}}
\hspace{5.em} \Muserfunction{Id}[b,\$1[6]]\ \Muserfunction{Id}[c,\$3[6]]+  \\
\noalign{\vspace{0.555556ex}}
\hspace{4.em} 5040\ \Muserfunction{Id}[a,\$1[6]]\   \\
\noalign{\vspace{0.555556ex}}
\hspace{5.em} \Muserfunction{Id}[b,\$2[6]]\ \Muserfunction{Id}[c,\$3[6]])\   \\
\noalign{\vspace{0.555556ex}}
\hspace{2.em} \Muserfunction{SB}[\$1[6],\$2[6],\$3[6]]\\
\MathEnd{MathArray}}

\end{CellGroup}
\mathin
In\ the\ previous\ input\ command\ line\ the\ \%\ symbol\ stands\ 

for\ "the\ last\ output".\ Similarly,\ \%\%\ stands\ for\ the\ 

output\ before\ the\ last\ one.\ Better\ way\ is\ to\ use\ the\ 

output\ name\ directly,\ like\ in\ Out[17]//.Rule1.\

The\ command\ //.Rule1\ means:\ apply\ Rule1\ as\ many\ times\ 

as\ it\ is\ applicable.\ If\ you\ want\ to\ apply\ it\ only\ once,\ 

the\ appropriate\ command\ is\ /.Rule1.\

What\ we\ have\ obtained\ in\ the\ last\ output\ is\ the\ correct\ 

answer\ in\ an\ unwanted\ form.\ To\ get\ the\ final\ answer\ we\ 

type:
\endmathin
\begin{CellGroup}
\dispSFinmath{\Mfunction{ExpandAll}[\%]}
\dispSFoutmath{48\ \Muserfunction{SB}[a,b,c]}

\end{CellGroup}
\mathin
When\ we\ put\ all\ this\ together,\ we\ see\ that\ the\ optimal\ 

command\ should\ have\ been:
\endmathin
\begin{CellGroup}
\dispSFinmath{\MathBegin{MathArray}{l}
\Mvariable{ExpandAll}[  \\
\noalign{\vspace{0.555556ex}}
\hspace{1.em} \Mfunction{ExpandAll}[\Muserfunction{SB}[e,f,g]**S[a,b,c]]**  \\
\noalign{\vspace{0.555556ex}}
\hspace{3.em} \Muserfunction{SB}[e,f,g]//.  \\
\noalign{\vspace{0.555556ex}}
\hspace{2.em} \Mvariable{Rule1}]\\
\MathEnd{MathArray}}
\dispSFoutmath{48\ \Muserfunction{SB}[a,b,c]}

\end{CellGroup}
\mathin
It\ is\ always\ better\ to\ use\ ExpandAll\ then\ Expand.\ 

Note\ also\ that\ Rule1\ has\ been\ used\ only\ once\ and\ at\ the\ 

end\ of\ the\ calculation!\ If\ you\ use\ it\ more\ than\ once\ in\ 

the\ same\ command\ line\ you\ will\ get\ wrong\ answer!\

Sometimes,\ you\ want\ to\ use\ only\ a\ portion\ of\ the\ intermediate\

result\ and\ multiply\ it\ with\ another\ sigma\ matrix\ -\ to\ check\ 

the\ result,\ for\ example.
\endmathin
\begin{CellGroup}
\dispSFinmath{\Mfunction{ExpandAll}[\Muserfunction{SB}[e,f,g]**S[a,b,c]]}
\dispSFoutmath{\MathBegin{MathArray}[p]{l}
-(\Muserfunction{Eta}[a,g]\ \Muserfunction{Eta}[b,f]\ \Muserfunction{Eta}[c,e])+  \\
\noalign{\vspace{0.555556ex}}
\hspace{1.em} \Muserfunction{Eta}[a,f]\ \Muserfunction{Eta}[b,g]\ \Muserfunction{Eta}[c,e]+  \\
\noalign{\vspace{0.555556ex}}
\hspace{1.em} \Muserfunction{Eta}[a,g]\ \Muserfunction{Eta}[b,e]\ \Muserfunction{Eta}[c,f]-  \\
\noalign{\vspace{0.555556ex}}
\hspace{1.em} \Muserfunction{Eta}[a,e]\ \Muserfunction{Eta}[b,g]\ \Muserfunction{Eta}[c,f]-  \\
\noalign{\vspace{0.555556ex}}
\hspace{1.em} \Muserfunction{Eta}[a,f]\ \Muserfunction{Eta}[b,e]\ \Muserfunction{Eta}[c,g]+  \\
\noalign{\vspace{0.555556ex}}
\hspace{1.em} \Muserfunction{Eta}[a,e]\ \Muserfunction{Eta}[b,f]\ \Muserfunction{Eta}[c,g]-  \\
\noalign{\vspace{0.555556ex}}
\hspace{1.em} \Muserfunction{Eta}[b,g]\ \Muserfunction{Eta}[c,f]\ S[a,e]+  \\
\noalign{\vspace{0.555556ex}}
\hspace{1.em} \Muserfunction{Eta}[b,f]\ \Muserfunction{Eta}[c,g]\ S[a,e]+  \\
\noalign{\vspace{0.555556ex}}
\hspace{1.em} \Muserfunction{Eta}[b,g]\ \Muserfunction{Eta}[c,e]\ S[a,f]-  \\
\noalign{\vspace{0.555556ex}}
\hspace{1.em} \Muserfunction{Eta}[b,e]\ \Muserfunction{Eta}[c,g]\ S[a,f]-  \\
\noalign{\vspace{0.555556ex}}
\hspace{1.em} \Muserfunction{Eta}[b,f]\ \Muserfunction{Eta}[c,e]\ S[a,g]+  \\
\noalign{\vspace{0.555556ex}}
\hspace{1.em} \Muserfunction{Eta}[b,e]\ \Muserfunction{Eta}[c,f]\ S[a,g]+  \\
\noalign{\vspace{0.555556ex}}
\hspace{1.em} \Muserfunction{Eta}[a,g]\ \Muserfunction{Eta}[c,f]\ S[b,e]-  \\
\noalign{\vspace{0.555556ex}}
\hspace{1.em} \Muserfunction{Eta}[a,f]\ \Muserfunction{Eta}[c,g]\ S[b,e]-  \\
\noalign{\vspace{0.555556ex}}
\hspace{1.em} \Muserfunction{Eta}[a,g]\ \Muserfunction{Eta}[c,e]\ S[b,f]+  \\
\noalign{\vspace{0.555556ex}}
\hspace{1.em} \Muserfunction{Eta}[a,e]\ \Muserfunction{Eta}[c,g]\ S[b,f]+  \\
\noalign{\vspace{0.555556ex}}
\hspace{1.em} \Muserfunction{Eta}[a,f]\ \Muserfunction{Eta}[c,e]\ S[b,g]-  \\
\noalign{\vspace{0.555556ex}}
\hspace{1.em} \Muserfunction{Eta}[a,e]\ \Muserfunction{Eta}[c,f]\ S[b,g]-  \\
\noalign{\vspace{0.555556ex}}
\hspace{1.em} \Muserfunction{Eta}[a,g]\ \Muserfunction{Eta}[b,f]\ S[c,e]+  \\
\noalign{\vspace{0.555556ex}}
\hspace{1.em} \Muserfunction{Eta}[a,f]\ \Muserfunction{Eta}[b,g]\ S[c,e]+  \\
\noalign{\vspace{0.555556ex}}
\hspace{1.em} \Muserfunction{Eta}[a,g]\ \Muserfunction{Eta}[b,e]\ S[c,f]-  \\
\noalign{\vspace{0.555556ex}}
\hspace{1.em} \Muserfunction{Eta}[a,e]\ \Muserfunction{Eta}[b,g]\ S[c,f]-  \\
\noalign{\vspace{0.555556ex}}
\hspace{1.em} \Muserfunction{Eta}[a,f]\ \Muserfunction{Eta}[b,e]\ S[c,g]+  \\
\noalign{\vspace{0.555556ex}}
\hspace{1.em} \Muserfunction{Eta}[a,e]\ \Muserfunction{Eta}[b,f]\ S[c,g]-  \\
\noalign{\vspace{0.555556ex}}
\hspace{1.em} \Muserfunction{Eta}[c,g]\ S[a,b,e,f]+  \\
\noalign{\vspace{0.555556ex}}
\hspace{1.em} \Muserfunction{Eta}[c,f]\ S[a,b,e,g]-  \\
\noalign{\vspace{0.555556ex}}
\hspace{1.em} \Muserfunction{Eta}[c,e]\ S[a,b,f,g]+  \\
\noalign{\vspace{0.555556ex}}
\hspace{1.em} \Muserfunction{Eta}[b,g]\ S[a,c,e,f]-  \\
\noalign{\vspace{0.555556ex}}
\hspace{1.em} \Muserfunction{Eta}[b,f]\ S[a,c,e,g]+  \\
\noalign{\vspace{0.555556ex}}
\hspace{1.em} \Muserfunction{Eta}[b,e]\ S[a,c,f,g]-  \\
\noalign{\vspace{0.555556ex}}
\hspace{1.em} \Muserfunction{Eta}[a,g]\ S[b,c,e,f]+  \\
\noalign{\vspace{0.555556ex}}
\hspace{1.em} \Muserfunction{Eta}[a,f]\ S[b,c,e,g]-  \\
\noalign{\vspace{0.555556ex}}
\hspace{1.em} \Muserfunction{Eta}[a,e]\ S[b,c,f,g]-  \\
\noalign{\vspace{0.555556ex}}
\hspace{1.em} S[a,b,c,e,f,g]\\
\MathEnd{MathArray}}

\end{CellGroup}
\begin{CellGroup}
\mathin
Eta\ symbol\ is\ the\ same\ as\ Id\ -\ depending\ whether\ both\ 

vector\ indices\ are\ down\ or\ up\ or\ not.\ Suppose\ now\ that\ 

you\ want\ to\ extract\ the\ last\ term\ S[a,b,c,e,f,g]\ and\ 

to\ multiply\ it\ with\ SB[e,f,g].\ You\ first\ click\ the\ mouse\

on\ the\ last\ output\ and\ press\ "Command-x",\ i.e.\ the\ 

Command\ key\ followed\ by\ x.\ This\ cuts\ the\ output.\ Put\ it\ 

back\ by\ pressing\ "Command-v".\ This\ will\ paste\ it\ back.\

Move\ the\ cursor\ beneath\ the\ last\ cell\ and\ press\ again

"Command-v".\ 
\endmathin
\dispSFoutmath{\MathBegin{MathArray}[p]{l}
-(\Muserfunction{Eta}[a,g]\ \Muserfunction{Eta}[b,f]\ \Muserfunction{Eta}[c,e])+  \\
\noalign{\vspace{0.555556ex}}
\hspace{1.em} \Muserfunction{Eta}[a,f]\ \Muserfunction{Eta}[b,g]\ \Muserfunction{Eta}[c,e]+  \\
\noalign{\vspace{0.555556ex}}
\hspace{1.em} \Muserfunction{Eta}[a,g]\ \Muserfunction{Eta}[b,e]\ \Muserfunction{Eta}[c,f]-  \\
\noalign{\vspace{0.555556ex}}
\hspace{1.em} \Muserfunction{Eta}[a,e]\ \Muserfunction{Eta}[b,g]\ \Muserfunction{Eta}[c,f]-  \\
\noalign{\vspace{0.555556ex}}
\hspace{1.em} \Muserfunction{Eta}[a,f]\ \Muserfunction{Eta}[b,e]\ \Muserfunction{Eta}[c,g]+  \\
\noalign{\vspace{0.555556ex}}
\hspace{1.em} \Muserfunction{Eta}[a,e]\ \Muserfunction{Eta}[b,f]\ \Muserfunction{Eta}[c,g]-  \\
\noalign{\vspace{0.555556ex}}
\hspace{1.em} \Muserfunction{Eta}[b,g]\ \Muserfunction{Eta}[c,f]\ S[a,e]+  \\
\noalign{\vspace{0.555556ex}}
\hspace{1.em} \Muserfunction{Eta}[b,f]\ \Muserfunction{Eta}[c,g]\ S[a,e]+  \\
\noalign{\vspace{0.555556ex}}
\hspace{1.em} \Muserfunction{Eta}[b,g]\ \Muserfunction{Eta}[c,e]\ S[a,f]-  \\
\noalign{\vspace{0.555556ex}}
\hspace{1.em} \Muserfunction{Eta}[b,e]\ \Muserfunction{Eta}[c,g]\ S[a,f]-  \\
\noalign{\vspace{0.555556ex}}
\hspace{1.em} \Muserfunction{Eta}[b,f]\ \Muserfunction{Eta}[c,e]\ S[a,g]+  \\
\noalign{\vspace{0.555556ex}}
\hspace{1.em} \Muserfunction{Eta}[b,e]\ \Muserfunction{Eta}[c,f]\ S[a,g]+  \\
\noalign{\vspace{0.555556ex}}
\hspace{1.em} \Muserfunction{Eta}[a,g]\ \Muserfunction{Eta}[c,f]\ S[b,e]-  \\
\noalign{\vspace{0.555556ex}}
\hspace{1.em} \Muserfunction{Eta}[a,f]\ \Muserfunction{Eta}[c,g]\ S[b,e]-  \\
\noalign{\vspace{0.555556ex}}
\hspace{1.em} \Muserfunction{Eta}[a,g]\ \Muserfunction{Eta}[c,e]\ S[b,f]+  \\
\noalign{\vspace{0.555556ex}}
\hspace{1.em} \Muserfunction{Eta}[a,e]\ \Muserfunction{Eta}[c,g]\ S[b,f]+  \\
\noalign{\vspace{0.555556ex}}
\hspace{1.em} \Muserfunction{Eta}[a,f]\ \Muserfunction{Eta}[c,e]\ S[b,g]-  \\
\noalign{\vspace{0.555556ex}}
\hspace{1.em} \Muserfunction{Eta}[a,e]\ \Muserfunction{Eta}[c,f]\ S[b,g]-  \\
\noalign{\vspace{0.555556ex}}
\hspace{1.em} \Muserfunction{Eta}[a,g]\ \Muserfunction{Eta}[b,f]\ S[c,e]+  \\
\noalign{\vspace{0.555556ex}}
\hspace{1.em} \Muserfunction{Eta}[a,f]\ \Muserfunction{Eta}[b,g]\ S[c,e]+  \\
\noalign{\vspace{0.555556ex}}
\hspace{1.em} \Muserfunction{Eta}[a,g]\ \Muserfunction{Eta}[b,e]\ S[c,f]-  \\
\noalign{\vspace{0.555556ex}}
\hspace{1.em} \Muserfunction{Eta}[a,e]\ \Muserfunction{Eta}[b,g]\ S[c,f]-  \\
\noalign{\vspace{0.555556ex}}
\hspace{1.em} \Muserfunction{Eta}[a,f]\ \Muserfunction{Eta}[b,e]\ S[c,g]+  \\
\noalign{\vspace{0.555556ex}}
\hspace{1.em} \Muserfunction{Eta}[a,e]\ \Muserfunction{Eta}[b,f]\ S[c,g]-  \\
\noalign{\vspace{0.555556ex}}
\hspace{1.em} \Muserfunction{Eta}[c,g]\ S[a,b,e,f]+  \\
\noalign{\vspace{0.555556ex}}
\hspace{1.em} \Muserfunction{Eta}[c,f]\ S[a,b,e,g]-  \\
\noalign{\vspace{0.555556ex}}
\hspace{1.em} \Muserfunction{Eta}[c,e]\ S[a,b,f,g]+  \\
\noalign{\vspace{0.555556ex}}
\hspace{1.em} \Muserfunction{Eta}[b,g]\ S[a,c,e,f]-  \\
\noalign{\vspace{0.555556ex}}
\hspace{1.em} \Muserfunction{Eta}[b,f]\ S[a,c,e,g]+  \\
\noalign{\vspace{0.555556ex}}
\hspace{1.em} \Muserfunction{Eta}[b,e]\ S[a,c,f,g]-  \\
\noalign{\vspace{0.555556ex}}
\hspace{1.em} \Muserfunction{Eta}[a,g]\ S[b,c,e,f]+  \\
\noalign{\vspace{0.555556ex}}
\hspace{1.em} \Muserfunction{Eta}[a,f]\ S[b,c,e,g]-  \\
\noalign{\vspace{0.555556ex}}
\hspace{1.em} \Muserfunction{Eta}[a,e]\ S[b,c,f,g]-  \\
\noalign{\vspace{0.555556ex}}
\hspace{1.em} S[a,b,c,e,f,g]\\
\MathEnd{MathArray}}

\end{CellGroup}
\mathin
Now\ you\ want\ this\ output\ to\ be\ turned\ into\ an\ input\ -

bold\ face.\ Click\ again\ on\ it\ and\ press\ "Command\ 9".\ This\ 

will\ turn\ it\ into\ a\ bold-faced\ form:

\endmathin
\dispSFinmath{\MathBegin{MathArray}[p]{l}
-(\Muserfunction{Eta}[a,g]\ \Muserfunction{Eta}[b,f]\ \Muserfunction{Eta}[c,e])+  \\
\noalign{\vspace{0.555556ex}}
\hspace{1.em} \Muserfunction{Eta}[a,f]\ \Muserfunction{Eta}[b,g]\ \Muserfunction{Eta}[c,e]+  \\
\noalign{\vspace{0.555556ex}}
\hspace{1.em} \Muserfunction{Eta}[a,g]\ \Muserfunction{Eta}[b,e]\ \Muserfunction{Eta}[c,f]-  \\
\noalign{\vspace{0.555556ex}}
\hspace{1.em} \Muserfunction{Eta}[a,e]\ \Muserfunction{Eta}[b,g]\ \Muserfunction{Eta}[c,f]-  \\
\noalign{\vspace{0.555556ex}}
\hspace{1.em} \Muserfunction{Eta}[a,f]\ \Muserfunction{Eta}[b,e]\ \Muserfunction{Eta}[c,g]+  \\
\noalign{\vspace{0.555556ex}}
\hspace{1.em} \Muserfunction{Eta}[a,e]\ \Muserfunction{Eta}[b,f]\ \Muserfunction{Eta}[c,g]-  \\
\noalign{\vspace{0.555556ex}}
\hspace{1.em} \Muserfunction{Eta}[b,g]\ \Muserfunction{Eta}[c,f]\ S[a,e]+  \\
\noalign{\vspace{0.555556ex}}
\hspace{1.em} \Muserfunction{Eta}[b,f]\ \Muserfunction{Eta}[c,g]\ S[a,e]+  \\
\noalign{\vspace{0.555556ex}}
\hspace{1.em} \Muserfunction{Eta}[b,g]\ \Muserfunction{Eta}[c,e]\ S[a,f]-  \\
\noalign{\vspace{0.555556ex}}
\hspace{1.em} \Muserfunction{Eta}[b,e]\ \Muserfunction{Eta}[c,g]\ S[a,f]-  \\
\noalign{\vspace{0.555556ex}}
\hspace{1.em} \Muserfunction{Eta}[b,f]\ \Muserfunction{Eta}[c,e]\ S[a,g]+  \\
\noalign{\vspace{0.555556ex}}
\hspace{1.em} \Muserfunction{Eta}[b,e]\ \Muserfunction{Eta}[c,f]\ S[a,g]+  \\
\noalign{\vspace{0.555556ex}}
\hspace{1.em} \Muserfunction{Eta}[a,g]\ \Muserfunction{Eta}[c,f]\ S[b,e]-  \\
\noalign{\vspace{0.555556ex}}
\hspace{1.em} \Muserfunction{Eta}[a,f]\ \Muserfunction{Eta}[c,g]\ S[b,e]-  \\
\noalign{\vspace{0.555556ex}}
\hspace{1.em} \Muserfunction{Eta}[a,g]\ \Muserfunction{Eta}[c,e]\ S[b,f]+  \\
\noalign{\vspace{0.555556ex}}
\hspace{1.em} \Muserfunction{Eta}[a,e]\ \Muserfunction{Eta}[c,g]\ S[b,f]+  \\
\noalign{\vspace{0.555556ex}}
\hspace{1.em} \Muserfunction{Eta}[a,f]\ \Muserfunction{Eta}[c,e]\ S[b,g]-  \\
\noalign{\vspace{0.555556ex}}
\hspace{1.em} \Muserfunction{Eta}[a,e]\ \Muserfunction{Eta}[c,f]\ S[b,g]-  \\
\noalign{\vspace{0.555556ex}}
\hspace{1.em} \Muserfunction{Eta}[a,g]\ \Muserfunction{Eta}[b,f]\ S[c,e]+  \\
\noalign{\vspace{0.555556ex}}
\hspace{1.em} \Muserfunction{Eta}[a,f]\ \Muserfunction{Eta}[b,g]\ S[c,e]+  \\
\noalign{\vspace{0.555556ex}}
\hspace{1.em} \Muserfunction{Eta}[a,g]\ \Muserfunction{Eta}[b,e]\ S[c,f]-  \\
\noalign{\vspace{0.555556ex}}
\hspace{1.em} \Muserfunction{Eta}[a,e]\ \Muserfunction{Eta}[b,g]\ S[c,f]-  \\
\noalign{\vspace{0.555556ex}}
\hspace{1.em} \Muserfunction{Eta}[a,f]\ \Muserfunction{Eta}[b,e]\ S[c,g]+  \\
\noalign{\vspace{0.555556ex}}
\hspace{1.em} \Muserfunction{Eta}[a,e]\ \Muserfunction{Eta}[b,f]\ S[c,g]-  \\
\noalign{\vspace{0.555556ex}}
\hspace{1.em} \Muserfunction{Eta}[c,g]\ S[a,b,e,f]+  \\
\noalign{\vspace{0.555556ex}}
\hspace{1.em} \Muserfunction{Eta}[c,f]\ S[a,b,e,g]-  \\
\noalign{\vspace{0.555556ex}}
\hspace{1.em} \Muserfunction{Eta}[c,e]\ S[a,b,f,g]+  \\
\noalign{\vspace{0.555556ex}}
\hspace{1.em} \Muserfunction{Eta}[b,g]\ S[a,c,e,f]-  \\
\noalign{\vspace{0.555556ex}}
\hspace{1.em} \Muserfunction{Eta}[b,f]\ S[a,c,e,g]+  \\
\noalign{\vspace{0.555556ex}}
\hspace{1.em} \Muserfunction{Eta}[b,e]\ S[a,c,f,g]-  \\
\noalign{\vspace{0.555556ex}}
\hspace{1.em} \Muserfunction{Eta}[a,g]\ S[b,c,e,f]+  \\
\noalign{\vspace{0.555556ex}}
\hspace{1.em} \Muserfunction{Eta}[a,f]\ S[b,c,e,g]-  \\
\noalign{\vspace{0.555556ex}}
\hspace{1.em} \Muserfunction{Eta}[a,e]\ S[b,c,f,g]-  \\
\noalign{\vspace{0.555556ex}}
\hspace{1.em} S[a,b,c,e,f,g]\\
\MathEnd{MathArray}}
\mathin
You\ can\ now\ darken\ the\ part\ which\ you\ want\ to\ throw\ out:\

Click\ the\ mouse\ at\ the\ beginning\ of\ the\ last\ input\ and\ 

drag\ it\ until\ the\ last\ term.\ Cut\ the\ darkened\ piece\ 

by\ "Command\ x".\ What\ is\ left\ is\ S[a,b,c,e,f,g].
\endmathin
\mathin
Below are two\ examples\ that correct Eqs(40) and (43) in\ the\

appendix\ of\ [6]:
\endmathin
\begin{CellGroup}
\dispSFinmath{\Mfunction{ExpandAll}[\Muserfunction{SB}[f,a,b]**S[f,c,d]]}
\dispSFoutmath{\MathBegin{MathArray}{l}
-8\ \Muserfunction{Eta}[a,d]\ \Muserfunction{Eta}[b,c]+  \\
\noalign{\vspace{0.555556ex}}
\hspace{1.em} 8\ \Muserfunction{Eta}[a,c]\ \Muserfunction{Eta}[b,d]-7\ \Muserfunction{Eta}[b,d]\ S[a,c]+  \\
\noalign{\vspace{0.555556ex}}
\hspace{1.em} 7\ \Muserfunction{Eta}[b,c]\ S[a,d]+7\ \Muserfunction{Eta}[a,d]\ S[b,c]-  \\
\noalign{\vspace{0.555556ex}}
\hspace{1.em} 7\ \Muserfunction{Eta}[a,c]\ S[b,d]-6\ S[a,b,c,d]\\
\MathEnd{MathArray}}

\end{CellGroup}
\begin{CellGroup}
\dispSFinmath{\Mfunction{ExpandAll}[\Muserfunction{SB}[a,b,c,d,e]**S[a,f,g]]}
\dispSFoutmath{\MathBegin{MathArray}[p]{l}
6\ \Muserfunction{Eta}[d,g]\ \Muserfunction{Eta}[e,f]\ S[b,c]-  \\
\noalign{\vspace{0.555556ex}}
\hspace{1.em} 6\ \Muserfunction{Eta}[d,f]\ \Muserfunction{Eta}[e,g]\ S[b,c]-  \\
\noalign{\vspace{0.555556ex}}
\hspace{1.em} 6\ \Muserfunction{Eta}[c,g]\ \Muserfunction{Eta}[e,f]\ S[b,d]+  \\
\noalign{\vspace{0.555556ex}}
\hspace{1.em} 6\ \Muserfunction{Eta}[c,f]\ \Muserfunction{Eta}[e,g]\ S[b,d]+  \\
\noalign{\vspace{0.555556ex}}
\hspace{1.em} 6\ \Muserfunction{Eta}[c,g]\ \Muserfunction{Eta}[d,f]\ S[b,e]-  \\
\noalign{\vspace{0.555556ex}}
\hspace{1.em} 6\ \Muserfunction{Eta}[c,f]\ \Muserfunction{Eta}[d,g]\ S[b,e]+  \\
\noalign{\vspace{0.555556ex}}
\hspace{1.em} 6\ \Muserfunction{Eta}[b,g]\ \Muserfunction{Eta}[e,f]\ S[c,d]-  \\
\noalign{\vspace{0.555556ex}}
\hspace{1.em} 6\ \Muserfunction{Eta}[b,f]\ \Muserfunction{Eta}[e,g]\ S[c,d]-  \\
\noalign{\vspace{0.555556ex}}
\hspace{1.em} 6\ \Muserfunction{Eta}[b,g]\ \Muserfunction{Eta}[d,f]\ S[c,e]+  \\
\noalign{\vspace{0.555556ex}}
\hspace{1.em} 6\ \Muserfunction{Eta}[b,f]\ \Muserfunction{Eta}[d,g]\ S[c,e]+  \\
\noalign{\vspace{0.555556ex}}
\hspace{1.em} 6\ \Muserfunction{Eta}[b,g]\ \Muserfunction{Eta}[c,f]\ S[d,e]-  \\
\noalign{\vspace{0.555556ex}}
\hspace{1.em} 6\ \Muserfunction{Eta}[b,f]\ \Muserfunction{Eta}[c,g]\ S[d,e]+  \\
\noalign{\vspace{0.555556ex}}
\hspace{1.em} 5\ \Muserfunction{Eta}[e,g]\ S[b,c,d,f]-  \\
\noalign{\vspace{0.555556ex}}
\hspace{1.em} 5\ \Muserfunction{Eta}[e,f]\ S[b,c,d,g]-  \\
\noalign{\vspace{0.555556ex}}
\hspace{1.em} 5\ \Muserfunction{Eta}[d,g]\ S[b,c,e,f]+  \\
\noalign{\vspace{0.555556ex}}
\hspace{1.em} 5\ \Muserfunction{Eta}[d,f]\ S[b,c,e,g]+  \\
\noalign{\vspace{0.555556ex}}
\hspace{1.em} 5\ \Muserfunction{Eta}[c,g]\ S[b,d,e,f]-  \\
\noalign{\vspace{0.555556ex}}
\hspace{1.em} 5\ \Muserfunction{Eta}[c,f]\ S[b,d,e,g]-  \\
\noalign{\vspace{0.555556ex}}
\hspace{1.em} 5\ \Muserfunction{Eta}[b,g]\ S[c,d,e,f]+  \\
\noalign{\vspace{0.555556ex}}
\hspace{1.em} 5\ \Muserfunction{Eta}[b,f]\ S[c,d,e,g]+  \\
\noalign{\vspace{0.555556ex}}
\hspace{1.em} 4\ S[b,c,d,e,f,g]\\
\MathEnd{MathArray}}

\end{CellGroup}

\section{Identities and Calculated Results}
\label{sec:3}

Before using the program to find new identities we use the definitions
from the first section to construct the following Fiertz identities.
Note that our program will not derive these identities since they are
not products of the elements via matrix multiplication.  
\begin{eqnarray}
&~&(\s^{\un a})_{(\a\b}(\s_{\un a})_{\g)\d}~=~0,\\
&~&(\s^{\un a})_{\a(\b}(\s_{\un a})_{\g)\d}~=-(\s^{\un a})_{\b\g}
(\s_{\un a})_{\a\d}, \\
&~&(\s^{\un a})^{\a(\b}(\s_{\un a})^{\g)\d}~=-(\s^{\un a})^{\b\g}
(\s_{\un a})^{\a\d},
\\
&~&(\s ^{[5]})_{\a\b}(\s _{[5]})_{\g\d}~=~0,\\
&~&(\s^{\un a \un b \un c})_{\a\b}(\s_{\un a \un b \un c})^{\g\d}=
-8\cdot 3!~\d_{[\a}^\g \d_{\b]}^\d,\\
&~&(\s^{\un a \un b \un c})_{\a\b}(\s_{\un a \un b \un c})_{\g\d}~=
-2\cdot 3!~
(\s^{\un a})_{\a[\g} (\s_{\un a})_{\d]\b},\\
&~&(\s^{\un a \un b \un c})^{\a\b}(\s_{\un a \un b \un c})^{\g\d}~=
-2\cdot 3!~
(\s^{\un a})^{\a[\g}(\s_{\un a})^{\d]\b},\\
&~&(\s^{[4]})_{\a}^{~\g}(\s_{[4]})_{\b}^{~\d}~=~4!\{
-2~\d_{\a}^\g \d_{\b}^\d +12~\d_{\b}^\g \d_{\a}^\d-
2~(\s ^{\un a})_{\a\b}(\s_{\un a})^{\g\d}\}, \\
&~&(\s^{[5]})_{\a\b}(\s_{[5]})^{\g\d}~=~5!\{-16~\d_{(\a}^\g \d_{\b)
}^\d-2~(\s ^{\un a})_{\a\b}(\s_{\un a})^{\g\d }\},\\
&~&(\s^{\un a \un b})_{\a}^{~\g}(\s_{{\un a \un b}})_{\b}^{~\d}~=~2~\{
-\d_{\a}^\g \d_{\b}^\d -4~\d_{\b}^\g \d_{\a}^\d-
2~(\s ^{\un a})_{\a\b}(\s _{\un a})^{\g\d}\},\\
&~&(\s^{\un a \un b})_{[\a}^{~\g}(\s_{{\un a \un b}})_{\b]}^{~\d}~=~
6~\d_{[\a}^\g \d_{\b]}^\d, \\
&~&(\s^{\un a \un b})_{(\a}^{~\g}(\s_{{\un a \un b}})_{\b)}^{~\d}~=
-10~\d_{(\a}^\g \d_{\b)}^\d -8~(\s ^{\un a})_{\a\b}(\s _{\un a})^{\g\d},\\
&~&[ 2~\d _{(\a}^\g \d _\b ^\d + (\s^{\un b})_{(\a\b |}(
\s_{\un b})^{\g\d}](
\s_{\un a})_{\e)\d}~=~0,\\
&~&\{ (\s_{[\un a})_{(\a |\g}(\s^{{\un e \un f \un g}})^{\g\d} + 
{1\over2}(\s^{{\un e \un f \un g}}
)_{(\a |\g} (\s_{[\un a})^{\g \d} \} (\s_{
\un b \un c \un d]})_{\b)\d}~=~48~(\s _{[\un a})_{\a
\b}~\d_{\un b}^{\un e} \d_{\un c}^{\un f}\d_{\un d]}^{\un g}  ,\\
&~&(\s_{\un a \un b \un c})_{(\a(\b}(\s^{\un c})_{\g)\d)}~=~-2~(\s _{[
\un a})_{\a\d}(\s_{\un b]})_{\b\g},\\
&~&(\s ^{\un a})_{(\a\b}(\s _{{\un a \un b}})_{\g)}^{~\d}~=~-(
\s_{\un b})_{(\a\b} \d_{\g)}^{\d}, \\
&~&(\s ^{\un a})^{(\a\b}(\s _{{\un a \un b}})^{~\g)}_{\d}~=~(\s_{\un b}
)^{(\a\b}\d^{\g)}_{\d},
\\
&~&(\s^{\un c})_{(\a\b}(\s_{{\un a \un b}})_{\g)}^{~\d}(
\s_{\un c})_{\d\e}
~=~-2(\s_{[\un a})_{(\a\b} (\s _{\un b]})_{\g)\e}.
\end{eqnarray}

There is also an interesting identity for which it is convenient
to first introduce two arbitrary three-forms in order to write
the identity in a simple manner,
\begin{eqnarray}
 (\s^{[3]}{}^{\un k \un l})_{(\a \b |} 
(\s^{[3'] }{}_{ \un k \un l})_{| \g \d ) } A_{[3]} B_{[3']}
 = ~  (\s^{\un a})_{(\a \b |}&&  (\s^{\un b})_{  |\g \d )}
 A_{\un a \un c \un d} B_{\un b}{}^{ \un c \un d} \nonumber \\ 
-~ 6 (\s^{\un a})_{(\a \b |} (\s^{\un b \un c \un d \un e \un f})_{
 | \g \d) } ( A_{\un a \un b \un c} B_{\un d \un e \un f} &+&   
B_{\un a \un b \un c} A_{\un d \un e \un f} )~~, 
\end{eqnarray}
valid for the arbitrary $A_{[3]} \equiv A_{\un a \un b 
\un c}$ and $B_{[3']} \equiv B_{\un d \un e \un f}$.

Finally, the following identities for manipulating the $\s$-matrices
are consequence of our definitions and are a direct test of the
reduction program, SigmaVector10D.m.  
\begin{eqnarray}
&~&(\s_{\un b \un c \un d})_{\a\b}(\s^{\un a \un b \un c})^{\b\g}~=~
72~\d_{\un d}^{\un a} \d_{\a}^\g~+~56(\s_{\un d} ^{~\un a})_{\a}^{~\g},\\
&~&(\s^{{\un a \un b}\un c \un d})_{\a}^{~\b} 
(\s_{\un b \un c \un d})_{\b\g}~=~
-7\cdot 8\cdot 9 ~(\s^{\un a})_{\a\g},\\
&~&(\s^{\un a \un b \un c})^{\a\b}(\s_{\un c})_{\b\g}~=~~8~(\s^{\un a
\un b})^{~\a}_{\g}, \\
&~&(\s^{\un a \un b \un c})_{\a\b}(\s_{\un c})^{\b\g}~=~-8~
(\s^{\un a \un b})_{\a}^{~\g}, \\
&~&(\s_{\un a})_{\a\b}(\s^{\un a \un b})^{~\b}_{\g}~=~-9~(\s^{b}
)_{\a\g}, \\
&~&(\s_{\un a})^{\a\b}(\s^{\un a \un b})^{~\g}_{\b}~=~~9~(
\s^{b})^{\a\g}, \\
&~&(\s^{{\un a \un b}cd})^{~\b}_{\a}(\s_{\un a})_{\b\g}~=~-7~(
\s^{\un b \un c \un d})_{\a\g},\\
&~&(\s^{{\un a \un b}\un c \un d})^{~\a}_{\b}(\s_{\un a}
)^{\b\g}~=~-7~(\s^{\un b \un c \un d})^{\a\g},\\
&~&(\s_{{\un a \un b}})_{\a}^{~\b}(\s^{\un c \un d})_{\b}^{~\g}~=~
-~\d_{[\un a}^{\un c} \d_{\un b]}^{\un d}
\d_{\a}^{\g}-\d_{[\un a}^{[\un c}(\s_{\un b]}^{~~\un d]})_{\a}^{~\g}+
(\s_{\un a \un b}^{~~\un c \un d})_{\a}^{~\g},\\
&~&(\s_{\un a \un b \un c})_{\a\b}(\s^{\un d \un e \un f})^{\b\g}~=~
~\d_{[\un a}^{\un d} \d_{\un b}^{\un e} \d_{\un c]}^{\un f}
\d_{\a}^{\g}+{1\over 2}\d_{[\un a}^{[\un d}\d_{\un b}^{\un e} 
(\s_{\un c]}^{~~\un f]})_{\a}^{~\g}-{1\over 4}\d_{[\un a}^{[\un d}(\s_{\un b
\un c]}^{~~\un e \un f]}
)_{\a}^{~\g}-(\s_{\un a \un b \un c}^{~~~\un d \un e \un f })_{\a}^{~\g},\\
&~&(\s^{\un a \un b})_{\a}^{~\b}(\s_{\un a \un b \un c})_{\b\g}~=~
-72~(\s_{\un c})_{\a\g} , \\
&~&(\s_{\un a \un b \un c})^{\a\b}(\s^{\un a \un b})_{\b}^{~\g}~=~
-72~(\s_{\un c})^{\a\g} ,\\
&~&(\s^{{\un a \un b}{\un c \un d \un e}})^{\a\b}(\s_{\un a \un 
b})_{\b}^{~\g}~=~
-6\cdot 7~(\s^{{\un c \un d \un e}})^{\a\g} ,\\
&~&(\s^{{\un a \un b}{\un c \un d \un e}})_{\a\b}(\s_{\un a 
\un b})_{\g}^{~\b}~=~
6\cdot 7~(\s^{{\un c \un d \un e}})_{\a\g} , \\
&~&(\s^{\un a \un b})_{\a}^{~\b}(\s_{\un a \un b}^{~~~\un c \un d}
)_{\b}^{~\g}~=~
-7\cdot 8~(\s^{\un c \un d})_{\a}^{~\g} ,\\
&~&(\s^{\un a \un b})_{\b}^{~\a}(\s_{\un a \un b}^{~~~\un c \un d}
)_{\g}^{~\b}~=~ -7\cdot 8~(\s^{\un c \un d})_{\g}^{~\a} ,\\
&~&(\s^{{\un a \un b}{\un c \un d \un e}})^{\a\b}(\s_{\un a \un b 
\un c})_{\b\g}~=~ -6\cdot 7\cdot
8~(\s^{\un d \un e})^{~\a}_{\g}, \\
&~&(\s^{{\un a \un b}{\un c \un d \un e}})^{\a\b}(\s_{b{\un c \un d 
\un e}})_{\b}^{~\g}~=~6\cdot 7 \cdot 8\cdot 9~(\s^{\un a})^{\a\g}, \\
&~&(\s^{{\un a \un b}{\un c \un d \un e}})^{\a\b}(\s_{\un a})_{\b\g}
~=~-6~(\s^{\un b{\un c \un d \un e}})^{~\a}_{\g}, \\
&~&(\s^{{\un a \un b}{\un c \un d \un e}})_{\a\b}(\s_{\un a})^{\b\g}~=~-6~
(\s^{\un b{\un c \un d \un e}}
)^{~\g}_{\a}, \\
&~&(\s^{\un a \un b}_{~~~\un c})^{\a\b}(\s_{{\un a \un b}\un d})_{\a\g}
~=~ 7 \cdot 8~(\s_{\un c \un d})^{~\b}
_{\g}-8\cdot 9 \h_{\un c \un d}\d_{\g}^{\b},\\
&~&(\s^{\un f}_{~~{\un a \un b}})^{\a\b}(\s_{\un f \un c \un d}
)_{\a\g}~=~ -6~(\s_{{\un a \un b}\un c \un d})^{~\b}_{\g}-
7 \h_{\un a[ \un c}(\s_{\un d] \un b})_{\g}^{~\b}-7\h_{\un b[ \un c}
(\s_{\un d] \un a})_{\g}^{~\b}+ 8\h_{\un a[ \un c}\h_{\un d] \un b}
\d_{\g}^{\b},\label{40} \\
&~&(\s^{{\un a \un b}{\un c \un d \un e}})^{\a\b}(\s_{{\un a \un b}\un f}
)_{\b\g}~=~ -5 \cdot 6~(\s_{\un f}^{~{\un c \un d \un e}})^
{~\a}_{\g}-{{6\cdot 7}\over 2}\d_{\un f}^{[c}(\s^{de]})_{\g}^{~\a},\\
&~&(\s^{\un a \un b \un c})^{\a\b}(\s_{[4]})_{\b}^{~\g}(\s_{\un a \un b}
)_{\g}^{~\d}~=~
4~(\s^{\un c})^{\a\b}(\s_{[4]})_{\b}^{~\d}~+~4~(\s_{[4]})_{\b}^{~\a}
(\s^{\un c})^{\b\d},
\\
&~&(\s^{{\un a \un b}{\un c \un d \un e}})^{\a\b}(\s_{\un a \un f
\un g})_{\b\g}~=~ 4~( \s_{\un f \un g}^{~~~\un b {\un c \un d \un e}}
)^{~\a}_{\g} + \frac{6}{2!}\d^{\un [d }_{\un [g } \d^{\un e}_{ \un
  f]} (\s^{\un b \un c]})^\a_\g + \frac{5}{3!} \d^{\un [e}_{\un [g} (\s^{\un b
  \un c \un d]}_{\un f]})^\a_\g
~ \label{43} \\
&~&(\s^{~{\un a \un b}}_{\un c})^{\a\b}(\s_{\un a \un d})_{\b}^{~\g}
(\s_{\un e \un b})_{\a}^{~\d}~=~ 3\cdot 19~(\s_{{\un c \un d \un e}})^{\g\d}
~~~~~~~~~~~~~~~~~~~~~~~~~~~~~~~~~~~if~~~\un c\ne \un d\ne \un e,\\
&~&(\s^{\un e}_{~{\un a \un b}})^{\a\b}(\s_{\un e \un c \un d}
)_{\a\g}~=~ 6~(\s_{{\un a \un b}\un c \un d})^{~\b}_{\g}
~~~~~~~~~~~~~~~~~~~~~~~~~~~~~~~~~~~~~~~~~~~if~~~[{\un a \un b}]\ne [\un c
\un d],\\
&~&(\s^{{\un a \un b}{\un c \un d \un e}})^{\a\b}(\s_{\un a \un f}
)_{\b}^{~\g} (\s_{\un b \un g})_{\a}^{~\d}~=~ 31
(\s_{\un f \un g}^{~~~{\un c \un d \un e}})^{\g\d}
~~~~~~~~~~~~~~~~~~~~~~~~~~~~~if~~~\un f \ne \un g \ne [{\un c \un d 
\un e}]\ne \un f,\\
&~&(\s^{\un a \un b}_{~~~{\un c \un d \un e}})^{\a\b}(\s_{\un a
\un f \un g})_{\b\g}~=~ 4~(\s_{~~{\un c \un d \un e}\un f \un g}^{\un b}
)^{~\a}_{\g}-5~\d_{[\un f}^{\un b}(\s_{\un g]{\un c \un d \un e}})_{\g}^{~\a}
~~~~~~~~~~~~~~~~~~~if~~~\un f\ne \un g\ne [{\un c \un d \un e}],\\
&~&(\s^{\un a \un b \un c})_{\a\g}(\s_{\un d \un e}
)_{\b}^{~\g}~=~ -~(\s^{\un a 
\un b \un c}_{~~~~\un d \un e})_{\a
\b} - {1 \over 2}\d_{[\un d}^{[\un a}(\s^{\un b \un c]}_{~~~~\un e]}
)_{\a\b}+\d_{\un d}^{[\un a}\d_{\un e}^{\un b} (\s^{\un c]})_{\a\b},\\
&~&(\s^{\un a \un b \un c})^{\a\g}(\s_{\un d \un e })_{\g}^{~\b}~=~
~(\s^{\un a \un b \un c}_{~~~~\un d \un e})^{\a\b}+
{1\over 2}\d_{[\un d}^{[\un a}(\s^{\un b \un c]}_{~~~~\un e]})^{\a\b}-
\d_{\un d}^{[\un a}\d_{\un e}^{\un b} (\s^{\un c]})^{\a\b},\\
&~&(\s_{{\un a \un b}{\un c \un d \un e}\un f})_{\b}^{~\a}(\s_{
\un g})^{\b\g}~=~ ~(\s_{{\un a \un b}{\un c \un d \un e}\un f \un g}
)^{\a\g}-
{1\over {5!}}\h_{g[\un a}(\s_{\un b{\un c \un d \un e}\un f]})^{\a\g},\\
&~&(\s^{\un a})_{\a\b}(\s_{\un b \un c \un d})^{\b\g}(\s^{\un e}
)_{\g\d}~=~- \h^{\un a \un e}(\s_{\un b \un c \un d})
_{\a\d}-{1\over 2}\d_{[\un b}^{[\un a}(\s_{\un c \un d]}^{~~~~\un e]}
)_{\a\d}+(\s^{\un a \un e}_{~~~\un b \un c \un d})_{\a\d}+
\d_{[\un b}^{\un a}\d_{\un c}^{\un e}(\s_{\un d]})_{\a\d},\\
&~&(\s_{[3]})^{\a\b}(\s_{\un a \un b})_{\b}^{~\g}(\s_{\un c \un d}
)_{\a}^{~\d}~=~{1\over {3!}} \e_{{\un a \un b}\un c \un d
[3]\bar{[3]}}(\s^{\bar{[3]}})^{\g\d}~~~~~~~~~~~~~~
~~~~~~~~if ~~~~[3]\ne [\un a, \un b, \un c,\un d],\\
&~&(\s^{\un d})_{(\a\b}(\s_{{\un a \un b}\un c \un d})_{\g)}^{~\d}
~={1\over 2}(\s_{[\un a})_{(\a\b}(\s_{\un b \un c]})_{\g)}^{~\d},\\
&~&(\s_{[\un a})_{(\a\b}(\s_{\un b \un c]})_{\g)}^{~\d}~=~-(\s_{
\un d})_{(\a\b}(\s^{\un d})^{\d\e}(\s_{\un a \un b \un c})_{\g)\e},\\
&~&(\s^{\un a \un b})_{\a}^{~\b}(\s_{\un c})_{\b\g}(\s_{\un a \un b}
)_{\d}^{~\g}~=~54~(\s_{\un c})_{\a\d},\\
&~&(\s^{\un a \un b})_{\a}^{~\b}(\s^{\un c \un d})_{\b}^{~\g}
(\s_{\un a \un b})_{\g}^{~\d}~=~-26~(\s^{\un c \un d})_{\a}^{~\d},\\
&~&(\s^{\un a \un b \un c})_{\a\g}(\s_{\un d \un c})_{\b}^{~\g}~=~7~(
\s_{\un d}^{~~{\un a \un b}})_{\a\b}-8~\d_{\un d}^{[\un a}(\s^{\un b]}
)_{\a\b},\\
&~&(\s_{\un a \un c})_{\b}^{~\a}(\s^{\un a \un b})_{\g}^{~\b}~
=~-8~(\s^{\un b}_{~~\un c})_{\g}^{~\a}-
9~\d_{\un c}^{\un b}\d_{\g}^{\a},\\
&~&(\s_{\un a \un b})_{\g}^{~\b}(\s^{[4]})_{\b}^{~\d}(\s^{\un b})_{\d\a}~=
~2~(\s_{\un a})_{\g\b}
(\s^{[4]})_{\a}^{~\b}-(\s^{[4]})_{\g}^{~\b}(\s_{\un a})_{\b\a},\\
&~&(\s^{\un b \un c})_{\g}^{~\b}(\s^{[4]})_{\b}^{~\d}(\s_{\un a \un b
\un c})_{\d\a}~=~2~(\s_{\un a})
_{\a\b}(\s^{[4]})_{\g}^{~\b}+4~(\s_{\un a})_{\g\b}(\s^{[4]})_{\a}^{~\b},\\
&~&(\s^{{\un a \un b}\un c \un d})_{\g}^{~\b}(\s^{[4]})_{\b}^{~\d}(\s_{
\un b \un c \un d})_{\d\a}~=~36~(\s^{\un a})_{\g\b}(\s^{[4]}
)_{\a}^{~\b}-6~(\s^{\un a})_{\a\b}(\s^{[4]})_{\g}^{~\b},
\\
&~&(\s_{\un a})_{\a\b}(\s^{\un a})^{\b\g}~=~-10~\d_{\a}^{\g},\\
&~&(\s_{\un a \un b})_{\a}^{~\b}(\s^{\un a \un b})_{\b}^{~\g}~=~-10\cdot 9
~\d_{\a}^{\g},\\
&~&(\s_{\un a \un b \un c})_{\a\b}(\s^{\un a \un b \un c})^{\b\g}~=~
10\cdot 9\cdot 8~\d_{\a}^{\g},
\\
&~&(\s_{{\un a \un b}\un c \un d})_{\a}^{~\b}(\s^{{\un a \un b}
\un c \un d})_{\b}^{~\g}
~=~10\cdot 9\cdot 8\cdot7~\d_{\a}^{\g},\\
&~&(\s_{{\un a \un b}{\un c \un d \un e}})_{\a\b}(\s^{{\un a \un b}
{\un c \un d \un e}})^{\b\g}~=~10
\cdot 9\cdot 8\cdot 7\cdot 6~\d_{\a}^{\g},\\
&~&(\s^{\un f})_{\a\b}(\s_{\un a})^{\b\g}(\s_{\un f})_{\g\d}~=~
8~(\s_{\un a})_{\a\d}, \\
%% $$\eqalign{
&~&(\s^{\un f})^{\a\b}(\s_{\un a})_{\b\g}(\s_{\un f})^{\g\d}~=~
8~(\s_{\un a})^{\a\d},\\
&~&(\s^{\un f})_{\a\b}(\s_{\un a \un b})^{~\b}_{\g}(\s_{\un f})^{\g\d}~=~
6~(\s_{\un a \un b})_{\a}^{~\d},\\
&~&(\s^{\un f})_{\a\b}(\s_{\un a \un b \un c})^{\b\g}(\s_{\un f}
)_{\g\d}~=~
4~(\s_{\un a \un b \un c})_{\a\d},\\
&~&(\s^{\un f})^{\a\b}(\s_{\un a \un b \un c})_{\b\g}(\s_{\un f}
)^{\g\d}~=~
4~(\s_{\un a \un b \un c})^{\a\d},\\
&~&(\s^{\un f})_{\a\b}(\s_{{\un a \un b}cd})^{~\b}_{\g}(\s_{\un f})^
{\g\d}~=~-2~(\s_{{\un a \un b}\un c \un d})_{\a}^{~\d},\\
&~&(\s^{\un f})_{\a\b}(\s_{[5]})^{\b\g}(\s_{\un f})_{\g\d}~=~0
,\\
&~&(\s^{{\un f}_{1}{\un f}_{2}})_{\a}^{~\b}(\s_{\un a})_{\b\g}(\s_{
{\un f}_{1}{\un f}_{2}})^{~\g}_{\d}~=~54~(\s_{\un a})_{\a\d},\\
&~&(\s^{{\un f}_{1}{\un f}_{2}})_{\a}^{~\b}(\s_{\un a})^{\a\g}
(\s_{{\un f}_{1}{\un f}_{2}})
^{~\d}_{\g}~=~54~(\s_{\un a})^{\b\d},\\
&~&(\s^{{\un f}_{1}{\un f}_{2}})_{\a}^{~\b}(\s_{\un a \un b})^{~\a}_{\g}
(\s_{{\un f}_{1}{\un f}_{2}})^{~\g}_{\d}~=~-26~(\s_{\un a \un b})^{~\b}_{\d}
,\\
&~&(\s^{{\un f}_{1}{\un f}_{2}})_{\a}^{~\b}(\s_{\un a \un b \un c}
)_{\b\g}(\s_{{\un f}_{1}{\un f}_{2}})^{~\g}_{\d}~=~6~(\s_{\un a \un b 
\un c})_{\a\d},\\
&~&(\s^{{\un f}_{1}{\un f}_{2}})_{\a}^{~\b}(\s_{\un a \un b \un c}
)^{\a\g}(\s_{{\un f}_{1}{\un f}_{2}})^{~\d}_{\g}~=~6~(\s_{\un a \un b 
\un c})^{\b\d},\\
&~&(\s^{{\un f}_{1}{\un f}_{2}})_{\a}^{~\b}(\s_{{\un a \un b}\un c \un d}
)^{~\a}_{\g}(\s_{{\un f}_{1}{\un f}_{2}})^{~\g}_{\d}~=~6~(\s_{{\un a \un b}
\un c \un d})^{~\b}_{\d},\\
&~&(\s^{{\un f}_{1}{\un f}_{2}})_{\a}^{~\b}(\s_{{\un a \un b}{\un c \un d 
\un e}})_{\b \g}
(\s_{{\un f}_{1}{\un f}_{2}})^{~\g}_{\d}~=~-10~(\s_{{\un a \un b}{\un c 
\un d \un e}})_{\a\d},\\
&~&(\s^{{\un f}_{1}{\un f}_{2}})_{\a}^{~\b}(\s_{{\un a \un b}{\un c \un d 
\un e}})^{\a \d}
(\s_{{\un f}_{1}{\un f}_{2}})^{~\g}_{\d}~=~-10~(\s_{{\un a \un b}{\un c \un d 
\un e}})^{\b\g},\\
&~&(\s^{{\un f}_{1}{\un f}_{2}{\un f}_{3}})_{\a\b}(\s_{\un a})^{\b\g}
(\s_{{\un f}_{1}{\un f}_{2}
{\un f}_{3}})_{\g \d}~=~-8\cdot 36~(\s_{\un a})_{\a\d},\\
&~&(\s^{{\un f}_{1}{\un f}_{2}{\un f}_{3}})^{\a\b}(\s_{\un a})_{\b\g}
(\s_{{\un f}_{1}{\un f}_{2}{\un f}_{3}}
)^{\g\d}~=~-8\cdot 36~(\s_{\un a})^{\a\d},\\
&~&(\s^{{\un f}_{1}{\un f}_{2}{\un f}_{3}})_{\a\b}(\s_{\un a \un b}
)^{~\b}_{\g}(\s_{{\un f}_{1}
{\un f}_{2}{\un f}_{3}})^{\g\d}~=~-48~(\s_{\un a \un b})_{\a}^{~\d},\\
&~&(\s^{{\un f}_{1}{\un f}_{2}{\un f}_{3}})_{\a\b}(\s_{\un a \un b 
\un c})^{\b\g} (\s_{{\un f}_{1}{\un f}_{2}{\un f}_{3}}
)_{\g\d}~=~ 48~(\s_{\un a \un b \un c})_{\a\d},\\
&~&(\s^{{\un f}_{1}{\un f}_{2}{\un f}_{3}})^{\a\b}(\s_{\un a \un b 
\un c})_{\b\g} (\s_{{\un f}_{1}{\un f}_{2}{\un f}_{3}}
)^{\g\d}~=~48~(\s_{\un a \un b \un c})^{\a\d},\\
&~&(\s^{{\un f}_{1}{\un f}_{2}{\un f}_{3}})_{\a\b}(\s_{{\un a \un b}\un c
\un d}
)^{~\b}_{\g}(\s_{{\un f}_{1}{\un f}_{2}
{\un f}_{3}})^{\g\d}~=~-48~(\s_{{\un a \un b}\un c \un d })_{\a}^{~\d},\\
&~&(\s^{{\un f}_{1}{\un f}_{2}{\un f}_{3}})_{\a\b}(\s_{[5]})^{\b\g}(\s_{
{\un f}_{1}{\un f}_{2}{\un f}_{3}}
)_{\g \d}~=~0 ,\\
&~&(\s^{{\un f}_{1}{\un f}_{2}{\un f}_{3}{\un f}_{4}})_{\a}^{~\b}(
\s_{\un a})_{\b\g}(\s_{{\un f}_{1}{\un f}_{2}
{\un f}_{3}{\un f}_{4}})^{~\g}_{\d}~=~24\cdot 42~(\s_{\un a})_{\a\d},\\
&~&(\s^{{\un f}_{1}{\un f}_{2}{\un f}_{3}{\un f}_{4}})_{\a}^{~\b}(
\s_{\un a})^{\a\g}(\s_{{\un f}_{1}{\un f}_{2}
{\un f}_{3}{\un f}_{4}})^{~\g}_{\d}~=~24\cdot 42~(\s_{\un a})^{\b\d},\\
&~&(\s^{{\un f}_{1}{\un f}_{2}{\un f}_{3}{\un f}_{4}})_{\a}^{~\b}(\s_{
\un a \un b})^{~\a}_{\g} (\s_{{\un f}_{1}
{\un f}_{2}{\un f}_{3}{\un f}_{4}})^{~\g}_{\d}~=~-24\cdot 14~(
\s_{\un a \un b})^{~\b}_{\d},\\
&~&(\s^{{\un f}_{1}{\un f}_{2}{\un f}_{3}{\un f}_{4}})_{\a}^{~\b}(
\s_{\un a \un b \un c})_{\b\g}
(\s_{{\un f}_{1}{\un f}_{2}{\un f}_{3}{\un f}_{4}})^{~\g}_{\d}~=~-24\cdot 
14~(\s_{\un a \un b \un c})_{\a\d}, \\
%% $$\eqalign{
&~&(\s^{{\un f}_{1}{\un f}_{2}{\un f}_{3}{\un f}_{4}})_{\a}^{~\b}
(\s_{\un a \un b \un c})^{\a\g}
(\s_{{\un f}_{1}{\un f}_{2}{\un f}_{3}{\un f}_{4}})_{\g}^{~\d} ~=~-24 
\cdot 14~(\s_{\un a \un b \un c} )^{\b\d} ,\\
&~&(\s^{{\un f}_{1}{\un f}_{2}{\un f}_{3}{\un f}_{4}})_{\a}^{~\b}(\s_{
{\un a \un b}\un c \un d })^{~\a}_{\g}
(\s_{{\un f}_{1}{\un f}_{2}{\un f}_{3}{\un f}_{4}})^{~\g}_{\d}~=~48~(\s_{
{\un a \un b}\un c \un d})^{~\b}_{\d}  ,\\
&~&(\s^{{\un f}_{1}{\un f}_{2}{\un f}_{3}{\un f}_{4}})_{\a}^{~\b}(\s_{
{\un a \un b}{\un c \un d \un e}})_{\b \g}
(\s_{{\un f}_{1}{\un f}_{2}{\un f}_{3}{\un f}_{4}})^{~\g}_{\d}~=~240~(\s_{
{\un a \un b}{\un c \un d \un e}})_{\a\d},\\
&~&(\s^{{\un f}_{1}{\un f}_{2}{\un f}_{3}{\un f}_{4}})_{\a}^{~\b}(\s_{
{\un a \un b}{\un c \un d \un e}})^{\a\d}
(\s_{{\un f}_{1}{\un f}_{2}{\un f}_{3}{\un f}_{4}})^{~\g}_{\d}~=~240~
(\s_{{\un a \un b}{
\un c \un d \un e} })^{\b\g}.
\end{eqnarray}
 
\section{The Program}
\label{sec:4}

%% 
%% This is a LaTeX document generated by automatic conversion
%% from a 
%% Mathematica notebook using Mathematica 3.0.
%% 
%% This document uses special macros defined in the style file
%% notebook.sty.
%% To run the document, you must put this style file in a
%% directory on
%% the path searched by LaTeX.  The style file is in the
%% IncludeFiles
%% subdirectory of the SystemFiles directory in your
%% Mathematica distribution.
%% 

%\def\CellGroup{\bgroup}
%\def\endCellGroup{\egroup}

%\documentstyle[notebook]{article}

%\begin{document}

%\mathin
(*BeginPackage["SigmaVector`"]*)

Print["This\ is\ SigmaVector\ program.\ It\ reduces\ the\ product\ of\ sigma\ matrices.\ 

Spinor\ indices\ of\ the\ chiral\ sigma\ matrices\ analyzed\ here\ are\ supressed.

Sigma\ matrices\ with\ spinor\ indices\ down\ are\ denoted\ by\ S,\ as\ well\ as\ those\ 

matrices\ which\ have\ one\ spinor\ index\ down\ and\ other\ up.\ If\ both\ spinor\ indices\ 

are\ up,\ the\ letter\ SB\ is\ used.\ 

The\ program\ handles\ all\ contraction\ possibilities\ of\ spinor\ indices,\ except\ when\ 

two\ sigma\ matrices,\ both\ with\ even\ number\ of\ vector\ indices\ are\ multiplied\ in\ such\ 

a\ way\ that\ the\ lower\ index\ of\ the\ first\ sigma\ matrix\ is\ summed\ with\ the\ upper\ index\ 

of\ the\ second\ matrix.\ In\ this\ case\ the\ order\ of\ sigma\ matrices\ should\ be\ reversed.

Here\ is\ an\ example:

$(\backslash s\RawWedge \{e,f\})\_\{\backslash a\}\RawWedge \{\backslash b\}*(\backslash s\_\{a,b,c\})\RawWedge \{\backslash a\backslash g\}*(\backslash s\_\{e,f\})\_\{\backslash g\}\RawWedge \{\backslash d\}$

should\ be\ represented\ in\ the\ symbolic\ form\ as:

$ExpandAll[ExpandAll[S[e,f]**SB[a,b,c]]**S[e,f]//.Rule1],$

where\ Rule1\ is\ a\ rule\ which\ transforms\ sigma\ matrices\ with\ 6\ or\ more\ vector\ indices\ 

in\ appropriate\ sigma\ matrices\ with\ 4\ or\ less\ vector\ indices\ and\ contracts\ corresponding\ 

epsilon\ symbols."]

(*\ These\ next\ rules\ assumed\ repeated\ indices\ for\ implied\ sum\ *)

SetAttributes[Id,Orderless]

Ten=10

SetAttributes[Eta,Orderless]

Eta/:\ Eta[a\_,b\_]\RawWedge 2:=\ Ten\ (*\ 10\ for\ our\ purposes\ *)

Eta/:\ Eta[a\_,a\_]:=\ Ten

Eta/:\ Eta[a\_,\ b\_]*S[d\_\_\_,a\_,\ c\_\_\_]\ :=\ S[d,b,\ c]

Eta/:\ S[d\_\_\_,a\_,\ c\_\_\_]*Eta[a\_,\ b\_]\ :=\ S[d,b,\ c]

Eta/:\ Eta[a\_,\ b\_]*SB[d\_\_\_,a\_,\ c\_\_\_]\ :=\ SB[d,b,\ c]

Eta/:\ SB[d\_\_\_,a\_,\ c\_\_\_]*Eta[a\_,\ b\_]\ :=\ SB[d,b,\ c]

Eta/:\ Eta[a\_,\ b\_]*Epsilon[d\_\_\_,a\_,\ c\_\_\_]\ :=\ Epsilon[d,b,\ c]

Eta/:\ Epsilon[d\_\_\_,a\_,\ c\_\_\_]*Eta[a\_,\ b\_]\ :=\ Epsilon[d,b,\ c]

Eta/:\ Eta[a\_,\ b\_]*Eta[a\_,\ c\_]\ :=\ Eta[b,\ c]

Eta/:\ Eta[a\_,\ c\_]*Eta[a\_,\ b\_]\ :=\ Eta[b,\ c]

Eta/:\ NumberQ[Eta[x\_\_]]\ :=\ True

(*\ These\ rules\ define\ Ten\ as\ the\ dimension\ so\ that\ its\ 

dependence\ in\ the

calculations\ can\ be\ followed.\ *)

Unprotect[NumberQ]

NumberQ[Ten\RawWedge n\_Integer]:=True

NumberQ[Ten]\ :=True

NumberQ[1/four!]:=True\ \ \ \ \ \ (*\ used\ to\ denote\ 1/4!\ *)

NumberQ[1/three!]:=True\ \ \ \ \ \ (*\ used\ to\ denote\ 1/3!\ *)

Protect[NumberQ]

(*\ \ Set\ the\ tag\ parameter\ for\ implied\ sums\ here\ *)

epi\ =\ 1

reset:=(epi=1)

Par[exp\_\_]:=(exp)

(*\ This\ comes\ from\ Epsilon.m\ \ *)

(*\ Rules\ for\ Epsilon\ when\ Euclidean\ space\ metric\ is\ present\ *)

(*\ These\ Rules\ are\ true\ for\ any\ dimension.\ \ Just\ change\ Ten\ to

the\ appropriate\ dimension.\ \ See\ also\ Rule2.\ \ *)

Epsilon/:\ NumberQ[Epsilon[x\_\_]]:=True

Epsilon/:\ Epsilon[b\_\_]*Epsilon[b\_\_]\ :=\ Ten!/;\ Signature[\{b\}]=!=0

Epsilon/:\ Epsilon[b\_\_]\RawWedge 2\ :=\ Ten!

Unprotect[NonCommutativeMultiply]

(x\_?NumberQ\ a\_[i\_\_])**(y\_?NumberQ\ b\_[j\_\_]):=(x\ y)\ a[i]**b[j]

(a\_\ +b\_)**c\_\ :=\ a**c\ +b**c

a\_\ **(b\_\ +\ c\_)\ :=\ a**b\ +\ a**c

a\_\ **(x\_?NumberQ\ b\_):=\ x\ a**b

(x\_?NumberQ\ a\_)**b\_\ :=\ x\ a**b

b\_?NumberQ\ **\ a\_\ :=\ b\ a

a\_\ **\ b\_?NumberQ\ :=\ b\ a

((anything\_\_)*c\_)**d\_\ :=\ (anything)(c**d)

Protect[NonCommutativeMultiply]

(*\ Definition\ of\ an\ antisymmetrizer.\ \ *)

SetAttributes[AntiSymmetrize,HoldAll]

AntiSymmetrize/:\ AntiSymmetrize[\ k\_\_,exp\_]\ :=

\ \ \ Sum[Signature[Permutations[k][[\$i]]]*Signature[k]*

\ \ \ \ \ (exp\ /.\ Table[k[[\$j]]\ $->$\ Permutations[k][[\$i]][[\$j]],\ 

\ \ \ \ \ \{\$j,\ Length[k]\}]),

\ \ \ \ \{\$i,\ 1,\ Length[k]!\}]

(*\ These\ rules\ define\ S,SB\ and\ Epsilon\ as\ purely\ 

anti-symmetric\ tensors

for\ any\ rank.\ *)

S[x\_\_]\ :=\ Signature[\{x\}]\ S@@Sort[\{x\}]\ /;\ Not[OrderedQ[\{x\}]]

SB[x\_\_]\ :=\ Signature[\{x\}]\ SB@@Sort[\{x\}]\ /;\ Not[OrderedQ[\{x\}]]

Epsilon[x\_\_]\ :=\ Signature[\{x\}]\ Epsilon@@Sort[\{x\}]\ /;\ 

Not[OrderedQ[\{x\}]]

S[x\_\_]:=0\ /;\ Signature[\{x\}]==0

SB[x\_\_]:=0\ /;\ Signature[\{x\}]==0

Epsilon[x\_\_]:=0\ /;\ Signature[\{x\}]==0

(*\ Epsilon[1,2,3,4,5,6,7,8,9,10]\ =\ 1\ *)

(*\ \ This\ rule\ replaces\ 6,7,8,9,\ and\ 10\ forms\ with\ their\ duals\ *)

Rule1\ =\ \{

S[a\_,b\_,c\_,d\_,e\_,f\_]\ :$>$\ 

(\ Epsilon[a,b,c,d,e,f,\$1[epi],\$2[epi],\$3[epi],\$4[epi]]*

S[\$1[epi],\$2[epi],\$3[epi],\$4[epi++]])/24,

S[d[1],d[2],d[3],d[4],d[5],d[6]]:$>$

(Epsilon[d[1],d[2],d[3],d[4],d[5],d[6],\$1[epi],\$2[epi],\$3[epi],\$4[epi]]*

S[\$1[epi],\$2[epi],\$3[epi],\$4[epi++]])/24,

SB[a\_,b\_,c\_,d\_,e\_,f\_,g\_]\ :$>$\ 

(-\ Epsilon[a,b,c,d,e,f,g,\$1[epi],\$2[epi],\$3[epi]]*

SB[\$1[epi],\$2[epi],\$3[epi++]])/6,

SB[d[1],d[2],d[3],d[4],d[5],d[6],d[7]]:$>$

(-Epsilon[d[1],d[2],d[3],d[4],d[5],d[6],d[7],1[epi],\$2[epi],\$3[epi]]*

SB[\$1[epi],\$2[epi],\$3[epi++]])/6,

S[a\_,b\_,c\_,d\_,e\_,f\_,g\_]\ :$>$\ 

(-\ Epsilon[a,b,c,d,e,f,g,\$1[epi],\$2[epi],\$3[epi]]*

S[\$1[epi],\$2[epi],\$3[epi++]])/6,

S[d[1],d[2],d[3],d[4],d[5],d[6],d[7]]:$>$\ 

(-Epsilon[d[1],d[2],d[3],d[4],d[5],d[6],d[7],\$1[epi],\$2[epi],\$3[epi]]*

S[\$1[epi],\$2[epi],\$3[epi++]])/6,

\ 

S[a\_,b\_,c\_,d\_,e\_,f\_,g\_,h\_]\ :$>$\ 

(-Epsilon[a,b,c,d,e,f,g,h,\$1[epi],\$2[epi]]*

S[\$1[epi],\$2[epi++]])/2,

S[d[1],d[2],d[3],d[4],d[5],d[6],d[7],d[8]]\ :$>$

(-Epsilon[d[1],d[2],d[3],d[4],d[5],d[6],d[7],d[8],\$1[epi],\$2[epi]]*

S[\$1[epi],\$2[epi++]])/2,

SB[a\_,b\_,c\_,d\_,e\_,f\_,g\_,h\_,i\_]\ :$>$\ 

(Epsilon[a,b,c,d,e,f,g,h,i,\$1[epi]]*

SB[\$1[epi++]]),

SB[d[1],d[2],d[3],d[4],d[5],d[6],d[7],d[8],d[9]]\ :$>$

(Epsilon[d[1],d[2],d[3],d[4],d[5],d[6],d[7],d[8],d[9],\$1[epi]]*

SB[\$1[epi++]]),

S[a\_,b\_,c\_,d\_,e\_,f\_,g\_,h\_,i\_]\ :$>$\ 

(Epsilon[a,b,c,d,e,f,g,h,i,\$1[epi]]*

S[\$1[epi++]])

\}

(*\ This\ rule\ pulls\ out\ the\ coefficients\ of\ chiral\ matrices.\ *)

SetAttributes[Eta2,Orderless]

Eta2/:\ NumberQ[Eta2[x\_\_]]\ :=\ True

Rule2\ =\ \{

S[a\_]\ :$>$\ S[d[1]]\ *\ Eta2[d[1],a],

S[a\_,b\_]\ :$>$\ S[d[1],d[2]]\ *\ Eta2[d[1],a]\ Eta2[d[2],b],

S[a\_,b\_,c\_]\ :$>$\ S[d[1],d[2],d[3]]*Eta2[d[1],a]*\ Eta2[d[2],b]\ Eta2[d[3],c],

S[a\_,b\_,c\_,d1\_]\ :$>$\ S[d[1],d[2],d[3],d[4]]\ *

Eta2[d[1],a]*\ Eta2[d[2],b]\ Eta2[d[3],c]\ Eta2[d[4],d1],

S[a\_,b\_,c\_,d1\_,e\_]\ :$>$S[d[1],d[2],d[3],d[4],d[5]]\ *\ 

Eta2[d[1],a]*\ Eta2[d[2],b]\ Eta2[d[3],c]\ Eta2[d[4],d1]\ Eta2[d[5],e],

S[a\_,b\_,c\_,d1\_,e\_,f\_]\ :$>$S[d[1],d[2],d[3],d[4],d[5],d[6]]\ *\ 

Eta2[d[1],a]*\ Eta2[d[2],b]\ Eta2[d[3],c]\ Eta2[d[4],d1]\ Eta2[d[5],e]\ *

Eta2[f,d[6]],

S[a\_,b\_,c\_,d1\_,e\_,f\_,g\_]\ :$>$S[d[1],d[2],d[3],d[4],d[5],d[6],d[7]]\ *\ 

Eta2[d[1],a]*\ Eta2[d[2],b]\ Eta2[d[3],c]\ Eta2[d[4],d1]\ Eta2[d[5],e]\ *

Eta2[f,d[6]]\ Eta2[g,d[7]],

S[a\_,b\_,c\_,d1\_,e\_,f\_,g\_,h\_]\ :$>$S[d[1],d[2],d[3],d[4],d[5],d[6],d[7],d[8]]\ *\ 

Eta2[d[1],a]*\ Eta2[d[2],b]\ Eta2[d[3],c]\ Eta2[d[4],d1]\ Eta2[d[5],e]\ *

Eta2[f,d[6]]\ Eta2[g,d[7]]\ Eta2[h,d[8]],

S[a\_,b\_,c\_,d1\_,e\_,f\_,g\_,h\_,i\_]\ :$>$S[d[1],d[2],d[3],d[4],d[5],d[6],d[7],d[8],d[9]]\ *\ 

Eta2[d[1],a]*\ Eta2[d[2],b]\ Eta2[d[3],c]\ Eta2[d[4],d1]\ Eta2[d[5],e]\ *

Eta2[f,d[6]]\ Eta2[g,d[7]]\ Eta2[h,d[8]]\ Eta2[i,d[9]],

S[a\_,b\_,c\_,d1\_,e\_,f\_,g\_,h\_,i\_,j\_]\ :$>$S[d[1],d[2],d[3],d[4],d[5],d[6],d[7],d[8],d[9],d[10]]\ *\ 

Eta2[d[1],a]*\ Eta2[d[2],b]\ Eta2[d[3],c]\ Eta2[d[4],d1]\ Eta2[d[5],e]\ *

Eta2[f,d[6]]\ Eta2[g,d[7]]\ Eta2[h,d[8]]\ Eta2[i,d[9]]\ Eta2[j,d[10]],

SB[a\_]\ :$>$\ SB[d[1]]\ *\ Eta2[d[1],a],

SB[a\_,b\_,c\_]\ :$>$\ SB[d[1],d[2],d[3]]*Eta2[d[1],a]*\ Eta2[d[2],b]\ Eta2[d[3],c],

SB[a\_,b\_,c\_,d1\_,e\_]\ :$>$SB[d[1],d[2],d[3],d[4],d[5]]\ *\ 

Eta2[d[1],a]*\ Eta2[d[2],b]\ Eta2[d[3],c]\ Eta2[d[4],d1]\ Eta2[d[5],e],

SB[a\_,b\_,c\_,d1\_,e\_,f\_,g\_]\ :$>$SB[d[1],d[2],d[3],d[4],d[5],d[6],d[7]]\ *\ 

Eta2[d[1],a]*\ Eta2[d[2],b]\ Eta2[d[3],c]\ Eta2[d[4],d1]\ Eta2[d[5],e]\ *

Eta2[f,d[6]]\ Eta2[g,d[7]],

SB[a\_,b\_,c\_,d1\_,e\_,f\_,g\_,h\_,i\_]\ :$>$SB[d[1],d[2],d[3],d[4],d[5],d[6],d[7],d[8],d[9]]\ *\ 

Eta2[d[1],a]*\ Eta2[d[2],b]\ Eta2[d[3],c]\ Eta2[d[4],d1]\ Eta2[d[5],e]\ *

Eta2[f,d[6]]\ Eta2[g,d[7]]\ Eta2[h,d[8]]\ Eta2[i,d[9]]

\}

(*\ Use\ Recover\ \ to\ factor\ out\ the\ S's\ and\ SB's\ *)

Recover[exp\_\_]\ :=\ Collect[exp/.Rule2,\{\ S[d[1]],\ S[d[1],d[2]],S[d[1],d[2],d[3]],

S[d[1],d[2],d[3],d[4]],\ S[d[1],d[2],d[3],d[4],d[5]],

S[d[1],d[2],d[3],d[4],d[5],d[6]],

S[d[1],d[2],d[3],d[4],d[5],d[6],d[7]],

S[d[1],d[2],d[3],d[4],d[5],d[6],d[7],d[8]],

S[d[1],d[2],d[3],d[4],d[5],d[6],d[7],d[8],d[9]],

S[d[1],d[2],d[3],d[4],d[5],d[6],d[7],d[8],d[9],d[10]],

SB[d[1]],\ SB[d[1],d[2],d[3]],\ SB[d[1],d[2],d[3],d[4],d[5]],

SB[d[1],d[2],d[3],d[4],d[5],d[6],d[7]],

SB[d[1],d[2],d[3],d[4],d[5],d[6],d[7],d[8],d[9]]\}\ ]/.Rule3

Rule3=\{Eta2[f\_\_]:$>$\ Eta[f]\}

\ 

(*\ \ This\ rule\ is\ time\ consuming.\ *)\ 

Epsilon/:\ Epsilon[b\_\_]*Epsilon[c\_\_]\ :=\ 

        -Par[\ Signature[\ Join[Complement[\{b\},

        Intersection[\{b\},\{c\}]],

        Intersection[\{b\},\{c\}]]]*

        Signature[\ Join[Complement[\{c\},Intersection[\{b\},\{c\}]],

        Intersection[\{b\},\{c\}]]]*ReleaseHold[\ (

\ \ \ \ \       AntiSymmetrize[Complement[\{c\},\ Intersection[\{b\},\ \{c\}]],

\ \ \ \ \ \     Product[Eta[Complement[\{b\},\ Intersection[\{b\},\ \{c\}]][[\$i]],

\ \ \ \ \ \ \ \ Complement[\{c\},\ Intersection[\{b\},\ \{c\}]][[\$i]]],

\ \ \ \ \ \ \   \{\$i,\ Length[Complement[\{c\},\ Intersection[\{b\},\ 

\ \ \ \ \ \ \   \{c\}]]]\}]])]*(Length[

\ \ \ \ \ \ \ \ Intersection[\{b\},\ \{c\}]]!)]\ /;\ \{b\}=!=\{c\}\ \&\&\ Length[\{b\}]==Length[\{c\}]\ 

S/:\ S[i\_]**SB[j\_]\ :=\ -\ Eta[i,\ j]\ -\ S[i,\ j]

S/:\ S[i\_]**S[j\_,k\_]\ :=\ -S[k]\ Eta[i,\ j]\ +\ S[j]\ Eta[i,\ k]\ -\ 

S[i,\ j,\ k]

S/:\ S[j\_,k\_]**S[i\_]\ :=\ -S[k]\ Eta[i,\ j]\ +\ S[j]\ Eta[i,\ k]\ +\ 

S[i,\ j,\ k]

S/:\ S[j\_,k\_]**SB[i\_]\ :=\ Eta[i,j]\ SB[k]\ -\ Eta[i,k]\ SB[j]\ -

SB[i,j,k]\ 

S/:\ S[i\_]\ **\ SB[j\_,\ k\_,\ l\_]\ :=

\ \ -S[k,\ l]\ Eta[i,\ j]\ -\ S[l,j]\ Eta[i,\ k]\ -\ S[j,\ k]\ Eta[i,\ l]\ -\ 

\ \ S[i,\ j,\ k,\ l]

S/:\ S[j\_,k\_,l\_]**SB[i\_]\ :=\ -Eta[i,j]\ S[k,l]\ -\ Eta[i,k]\ S[l,j]\ -\ 

Eta[i,l]\ S[j,k]\ +\ S[i,j,k,l]

S/:\ S[i\_]**S[j\_,k\_,l\_,m\_]\ :=\ Eta[i,j]\ S[k,l,m]\ -\ 

Eta[i,k]\ S[l,m,j]\ +\ Eta[i,l]\ S[j,k,m]\ -

Eta[i,m]\ S[j,k,l]\ +\ S[i,j,k,l,m]

S/:\ S[j\_,k\_,l\_,m\_]**S[i\_]\ :=\ -Eta[i,j]\ S[k,l,m]\ +\ 

Eta[i,k]\ S[l,m,j]\ -\ Eta[i,l]\ S[j,k,m]\ +

Eta[i,m]\ S[j,k,l]\ +\ S[i,j,k,l,m]

S/:\ S[j\_,k\_,l\_,m\_]**SB[i\_]\ :=\ -Eta[i,j]\ SB[k,l,m]\ +\ 

Eta[i,k]\ SB[l,m,j]\ -\ Eta[i,l]\ SB[j,k,m]\ +

Eta[i,m]\ SB[j,k,l]\ +\ SB[i,j,k,l,m]

S/:\ S[i\_]**SB[j\_,k\_,l\_,m\_,n\_]\ :=\ -Eta[i,j]\ S[k,l,m,n]\ -\ 

Eta[i,k]\ S[l,m,n,j]\ -\ Eta[i,l]\ S[m,n,j,k]\ -\ 

Eta[i,m]\ S[n,j,k,l]\ -\ Eta[i,n]\ S[j,k,l,m]\ -\ S[i,j,k,l,m,n]\ 

S/:\ S[j\_,k\_,l\_,m\_,n\_]**SB[i\_]\ :=\ \ -Eta[i,j]\ S[k,l,m,n]\ -\ 

Eta[i,k]\ S[l,m,n,j]\ -\ Eta[i,l]\ S[m,n,j,k]\ -\ 

Eta[i,m]\ S[n,j,k,l]\ -\ Eta[i,n]\ S[j,k,l,m]\ +\ S[i,j,k,l,m,n]\ 

S/:\ S[i\_]**S[j\_,k\_,l\_,m\_,n\_,p\_]\ :=\ -\ S[i,j,k,l,m,n,p]\ +

(Epsilon[j,k,l,m,n,p,\$1[epi],\ \$2[epi],\ \$3[epi],\ \$4[epi]]*

S[i,\$1[epi],\ \$2[epi],\ \$3[epi],\ \$4[epi++]])/24

S/:\ S[j\_,k\_,l\_,m\_,n\_,p\_]**SB[i\_]\ :=\ \ SB[i,j,k,l,m,n,p]\ +

(Epsilon[j,k,l,m,n,p,\$1[epi],\ \$2[epi],\ \$3[epi],\ \$4[epi]]*

SB[i,\$1[epi],\ \$2[epi],\ \$3[epi],\ \$4[epi++]])/24

S/:\ S[j\_,k\_,l\_,m\_,n\_,p\_]**S[i\_]\ :=\ S[i,j,k,l,m,n,p]\ +

(Epsilon[j,k,l,m,n,p,\$1[epi],\ \$2[epi],\ \$3[epi],\ \$4[epi]]*

S[i,\$1[epi],\ \$2[epi],\ \$3[epi],\ \$4[epi++]])/24

S/:\ S[i\_]**SB[j\_,k\_,l\_,m\_,n\_,p\_,q\_]\ :=\ \ S[i,j,k,l,m,n,p,q]\ +

(Epsilon[j,k,l,m,n,p,q,\$1[epi],\ \$2[epi],\ \$3[epi]]*

S[i,\$1[epi],\ \$2[epi],\ \$3[epi++]])/6

S/:\ S[j\_,k\_,l\_,m\_,n\_,p\_,q\_]**SB[i\_]\ :=\ \ S[i,j,k,l,m,n,p,q]\ -

(Epsilon[j,k,l,m,n,p,q,\$1[epi],\ \$2[epi],\ \$3[epi]]*

S[i,\$1[epi],\ \$2[epi],\ \$3[epi++]])/6

S/:\ S[i\_]**S[j\_,k\_,l\_,m\_,n\_,o\_,p\_,q\_]\ :=\ \ 

S[i,j,k,l,m,n,o,p,q]+

(Epsilon[j,k,l,m,n,o,p,q,\$1[epi],\ \$2[epi]]*

S[i,\$1[epi],\ \$2[epi++]])/2

S/:\ S[j\_,k\_,l\_,m\_,n\_,o\_,p\_,q\_]**S[i\_]\ :=\ \ 

S[i,j,k,l,m,n,o,p,q]-

(Epsilon[j,k,l,m,n,o,p,q,\$1[epi],\ \$2[epi]]*

S[i,\$1[epi],\ \$2[epi++]])/2

S/:\ S[j\_,k\_,l\_,m\_,n\_,o\_,p\_,q\_]**SB[i\_]\ :=\ \ 

-\ SB[i,j,k,l,m,n,o,p,q]+

(Epsilon[j,k,l,m,n,o,p,q,\$1[epi],\ \$2[epi]]*

SB[i,\$1[epi],\ \$2[epi++]])/2

S/:\ S[i\_]**SB[j\_,k\_,l\_,m\_,n\_,o\_,p\_,q\_,r\_]\ :=\ \ 

Epsilon[i,j,k,l,m,n,o,p,q,r]\ -\ 

Epsilon[j,k,l,m,n,o,p,q,r,\$1[epi]]*S[i,\$1[epi++]]

S/:\ S[j\_,k\_,l\_,m\_,n\_,o\_,p\_,q\_,r\_]**SB[i\_]\ :=\ \ 

\ Epsilon[i,j,k,l,m,n,o,p,q,r]\ +\ 

Epsilon[j,k,l,m,n,o,p,q,r,\$1[epi]]*S[i,\$1[epi++]]

SB/:\ SB[i\_]\ **\ S[j\_]\ :=\ -Eta[i,\ j]\ +\ S[i,\ j]

SB/:\ SB[i\_]**S[j\_,k\_]\ :=\ Eta[i,j]\ SB[k]\ -\ Eta[i,k]\ SB[j]\ +\ 

SB[i,j,k]

SB/:\ SB[i\_]**S[j\_,k\_,l\_]\ :=\ Eta[i,j]\ S[k,l]\ +\ 

Eta[i,k]\ S[l,j]\ +\ Eta[i,l]\ S[j,k]\ -\ S[i,j,k,l]

SB/:\ SB[j\_,\ k\_,\ l\_]\ **\ S[i\_]\ :=

\ S[k,\ l]\ Eta[i,\ j]\ +\ S[l,j]\ Eta[i,\ k]\ +\ S[j,\ k]\ Eta[i,\ l]\ +\ 

\ S[i,\ j,\ k,\ l]

SB/:\ SB[i\_]**S[j\_,k\_,l\_,m\_]\ :=\ Eta[i,j]\ SB[k,l,m]\ -\ 

Eta[i,k]\ SB[l,m,j]\ +\ Eta[i,l]\ SB[j,k,m]\ -

Eta[i,m]\ SB[j,k,l]\ +\ SB[i,j,k,l,m]

SB/:\ SB[i\_]**S[j\_,k\_,l\_,m\_,n\_]\ :=\ \ -Eta[i,j]\ S[k,l,m,n]\ -\ 

Eta[i,k]\ S[l,m,n,j]\ -\ Eta[i,l]\ S[m,n,j,k]\ -\ 

Eta[i,m]\ S[n,j,k,l]\ -\ Eta[i,n]\ S[j,k,l,m]\ +\ S[i,j,k,l,m,n]\ 

SB/:\ SB[j\_,k\_,l\_,m\_,n\_]**S[i\_]\ :=\ -Eta[i,j]\ S[k,l,m,n]\ -\ 

Eta[i,k]\ S[l,m,n,j]\ -\ Eta[i,l]\ S[m,n,j,k]\ -\ 

Eta[i,m]\ S[n,j,k,l]\ -\ Eta[i,n]\ S[j,k,l,m]\ -\ S[i,j,k,l,m,n]\ 

SB/:\ SB[i\_]**S[j\_,k\_,l\_,m\_,n\_,p\_]\ :=\ -\ SB[i,j,k,l,m,n,p]\ +

(Epsilon[j,k,l,m,n,p,\$1[epi],\ \$2[epi],\ \$3[epi],\ \$4[epi]]*

SB[i,\$1[epi],\ \$2[epi],\ \$3[epi],\ \$4[epi++]])/24

SB/:\ SB[j\_,k\_,l\_,m\_,n\_,p\_,q\_]**S[i\_]\ :=\ -S[i,j,k,l,m,n,p,q]-

(Epsilon[j,k,l,m,n,p,q,\$1[epi],\ \$2[epi],\ \$3[epi]]*

S[i,\$1[epi],\ \$2[epi],\ \$3[epi++]])/6

SB/:\ SB[i\_]**S[j\_,k\_,l\_,m\_,n\_,p\_,q\_]\ :=\ -S[i,j,k,l,m,n,p,q]+

(Epsilon[j,k,l,m,n,p,q,\$1[epi],\ \$2[epi],\ \$3[epi]]*

S[i,\$1[epi],\ \$2[epi],\ \$3[epi++]])/6

SB/:\ SB[i\_]**S[j\_,k\_,l\_,m\_,n\_,o\_,p\_,q\_]\ :=\ \ 

-\ SB[i,j,k,l,m,n,o,p,q]-

(Epsilon[j,k,l,m,n,o,p,q,\$1[epi],\ \$2[epi]]*

SB[i,\$1[epi],\ \$2[epi++]])/2

SB/:\ SB[j\_,k\_,l\_,m\_,n\_,o\_,p\_,q\_,r\_]**S[i\_]\ :=\ \ 

Epsilon[i,j,k,l,m,n,o,p,q,r]\ -\ 

Epsilon[j,k,l,m,n,o,p,q,r,\$1[epi]]*S[i,\$1[epi++]]

SB/:\ SB[i\_]**S[j\_,k\_,l\_,m\_,n\_,o\_,p\_,q\_,r\_]\ :=\ \ 

\ Epsilon[i,j,k,l,m,n,o,p,q,r]\ +\ 

Epsilon[j,k,l,m,n,o,p,q,r,\$1[epi]]*S[i,\$1[epi++]]

S/:\ S[i\_,j\_]**S[k\_,l\_]\ :=\ -Eta[k,l]\ S[i,j]\ -

Expand[\ Par[S[i,j]**S[k]]**SB[l]]

S/:\ S[a\_,b\_,c\_,d\_]**S[k\_,l\_]\ :=\ 

-Eta[k,l]\ S[a,b,c,d]\ -

Expand[\ Par[S[a,b,c,d]**S[k]]**SB[l]]

S/:\ S[a\_,b\_,c\_,d\_,e\_,f\_]**S[k\_,l\_]\ :=\ 

-Eta[k,l]\ S[a,b,c,d,e,f]\ -

Expand[\ Par[S[a,b,c,d,e,f]**S[k]]**SB[l]]

S/:\ S[a\_,b\_,c\_,d\_,e\_,f\_,g\_,h\_]**S[k\_,l\_]\ :=\ 

-Eta[k,l]\ S[a,b,c,d,e,f,g,h]\ -

Expand[\ Par[S[a,b,c,d,e,f,g,h]**S[k]]**SB[l]]

S/:\ S[a\_,b\_,c\_]**S[k\_,l\_]\ :=\ 

Eta[k,l]\ S[a,b,c]\ +

Expand[\ Par[S[a,b,c]**SB[k]]**S[l]]

S/:\ S[a\_,b\_,c\_,d\_,e\_]**S[k\_,l\_]\ :=\ 

Eta[k,l]\ S[a,b,c,d,e]\ +

Expand[\ Par[S[a,b,c,d,e]**SB[k]]**S[l]]

S/:\ S[a\_,b\_,c\_,d\_,e\_,f\_,g\_]**S[k\_,l\_]\ :=\ 

Eta[k,l]\ S[a,b,c,d,e,f,g]\ +

Expand[\ Par[S[a,b,c,d,e,f,g]**SB[k]]**S[l]]

S/:\ S[a\_,b\_,c\_,d\_,e\_,f\_,g\_,h\_,m\_]**S[k\_,l\_]\ :=\ 

Eta[k,l]\ S[a,b,c,d,e,f,g,h,m]\ +

Expand[\ Par[S[a,b,c,d,e,f,g,h,m]**SB[k]]**S[l]]

S/:\ S[i\_,j\_]**S[k\_,l\_\_]\ :=\ -Eta[i,j]\ S[k,l]\ -

Expand[\ S[i]**Par[SB[j]**S[k,l]]]

S/:\ S[i\_,j\_]**SB[k\_,l\_\_]\ :=\ Eta[i,j]\ SB[k,l]\ +

Expand[\ SB[i]**Par[S[j]**SB[k,l]]]

SB/:\ SB[i\_,j\_\_]**S[k\_,l\_]\ :=\ -\ Eta[k,l]\ SB[i,j]\ -\ 

Expand[Par[SB[i,j]**S[k]]**SB[l]]

S/:\ S[i\_,j\_,k\_]**S[l\_,m\_,n\_\_]\ :=\ 

-\ Expand[S[i]**Par[SB[j]**Par[S[k]**S[l,m,n]]]]\ -\ 

Eta[i,j]\ S[k]**S[l,m,n]\ +\ Eta[i,k]\ S[j]**S[l,m,n]-

Eta[j,k]\ S[i]**S[l,m,n]\ \ \ 

S/:\ S[i\_,j\_,k\_]**SB[l\_,m\_,n\_\_]\ :=\ 

-\ Expand[S[i]**Par[SB[j]**

Par[S[k]**SB[l,m,n]]]]\ -\ 

Eta[i,j]\ S[k]**SB[l,m,n]\ +\ Eta[i,k]\ S[j]**SB[l,m,n]\ -\ 

Eta[j,k]\ S[i]**SB[l,m,n]

S/:\ S[i\_,j\_,k\_\_]**S[l\_,m\_,n\_]\ :=\ 

-\ Expand[Par[Par[S[i,j,k]**S[l]]**SB[m]]**

S[n]]\ -\ 

Eta[l,m]\ S[i,j,k]**S[n]\ +\ Eta[l,n]\ S[i,j,k]**S[m]\ -\ 

Eta[m,n]\ S[i,j,k]**S[l]

S/:\ S[a\_,b\_,c\_\_]**SB[i\_,j\_,k\_]\ :=\ 

-\ Expand[Par[Par[S[a,b,c]**SB[j]]**S[k]]**

SB[i]]\ -\ 

Eta[j,k]\ S[a,b,c]**SB[i]\ +\ Eta[i,j]\ S[a,b,c]**SB[k]\ -\ 

Eta[i,k]\ S[a,b,c]**SB[j]

S/:\ SB[a\_,b\_,c\_\_]**S[i\_,j\_,k\_]\ :=\ 

-\ Expand[Par[Par[SB[a,b,c]**S[i]]**SB[j]]**S[k]]\ -\ 

Eta[j,k]\ SB[a,b,c]**S[i]\ -\ Eta[i,j]\ SB[a,b,c]**S[k]\ +\ 

Eta[i,k]\ SB[a,b,c]**S[j]

SB/:\ SB[i\_,j\_,k\_]**S[l\_,m\_,n\_\_]\ :=\ 

-\ Expand[SB[j]**Par[S[k]**

Par[SB[i]**S[l,m,n]]]]\ +\ 

Eta[i,j]\ SB[k]**S[l,m,n]\ -\ Eta[i,k]\ SB[j]**S[l,m,n]\ -

Eta[j,k]\ SB[i]**S[l,m,n]

S/:\ S[a\_,b\_,c\_,d\_]**S[i\_,j\_,k\_,l\_\_]\ :=\ 

Expand[S[b,c,d]**Par[SB[a]**S[i,j,k,l]]]+

Eta[a,b]\ Par[S[c,d]**S[i,j,k,l]]-

Eta[a,c]\ Par[S[b,d]**S[i,j,k,l]]+

Eta[a,d]\ Par[S[b,c]**S[i,j,k,l]]

S/:\ S[a\_,b\_,c\_,d\_]**SB[i\_,j\_,k\_,l\_\_]\ :=\ 

-\ Expand[SB[a]**Par[S[b,c,d]**SB[i,j,k,l]]]+

Eta[a,b]\ Par[S[c,d]**SB[i,j,k,l]]-

Eta[a,c]\ Par[S[b,d]**SB[i,j,k,l]]+

Eta[a,d]\ Par[S[b,c]**SB[i,j,k,l]]

S/:\ S[a\_,b\_,c\_,d\_]**S[i\_,j\_,k\_,l\_]\ :=\ 

-\ Expand[Par[S[a,b,c,d]**S[i]]**SB[j,k,l]]-

Eta[i,j]\ Par[S[a,b,c,d]**S[k,l]]+

Eta[i,k]\ Par[S[a,b,c,d]**S[j,l]]-

Eta[i,l]\ Par[S[a,b,c,d]**S[j,k]]

S/:\ S[a\_,b\_,c\_,d\_,e\_,f\_]**S[i\_,j\_,k\_,l\_]\ :=\ 

-\ Expand[Par[S[a,b,c,d,e,f]**S[i]]**SB[j,k,l]]-

Eta[i,j]\ Par[S[a,b,c,d,e,f]**S[k,l]]+

Eta[i,k]\ Par[S[a,b,c,d,e,f]**S[j,l]]-

Eta[i,l]\ Par[S[a,b,c,d,e,f]**S[j,k]]

S/:\ S[a\_,b\_,c\_,d\_,e\_,f\_,g\_,h\_]**S[i\_,j\_,k\_,l\_]\ :=\ 

-\ Expand[Par[S[a,b,c,d,e,f,g,h]**S[i]]**SB[j,k,l]]-

Eta[i,j]\ Par[S[a,b,c,d,e,f,g,h]**S[k,l]]+

Eta[i,k]\ Par[S[a,b,c,d,e,f,g,h]**S[j,l]]-

Eta[i,l]\ Par[S[a,b,c,d,e,f,g,h]**S[j,k]]

S/:\ S[a\_,b\_,c\_,d\_,e\_]**S[i\_,j\_,k\_,l\_]\ :=\ 

-\ Expand[

Par[S[a,b,c,d,e]**SB[i]]**S[j,k,l]]+

Eta[i,j]\ Par[S[a,b,c,d,e]**S[k,l]]-

Eta[i,k]\ Par[S[a,b,c,d,e]**S[j,l]]+

Eta[i,l]\ Par[S[a,b,c,d,e]**S[j,k]]

S/:\ S[a\_,b\_,c\_,d\_,e\_,f\_,g\_]**S[i\_,j\_,k\_,l\_]\ :=\ 

-\ Expand[

Par[S[a,b,c,d,e,f,g]**SB[i]]**S[j,k,l]]+

Eta[i,j]\ Par[S[a,b,c,d,e,f,g]**S[k,l]]-

Eta[i,k]\ Par[S[a,b,c,d,e,f,g]**S[j,l]]+

Eta[i,l]\ Par[S[a,b,c,d,e,f,g]**S[j,k]]

S/:\ S[a\_,b\_,c\_,d\_,e\_,f\_,g\_,h\_,m\_]**S[i\_,j\_,k\_,l\_]\ :=\ 

-\ Expand[

Par[S[a,b,c,d,e,f,g,h,m]**SB[i]]**S[j,k,l]]+

Eta[i,j]\ Par[S[a,b,c,d,e,f,g,h,m]**S[k,l]]-

Eta[i,k]\ Par[S[a,b,c,d,e,f,g,h,m]**S[j,l]]+

Eta[i,l]\ Par[S[a,b,c,d,e,f,g,h,m]**S[j,k]]

S/:\ SB[a\_,b\_,c\_,d\_\_]**S[i\_,j\_,k\_,l\_]\ :=\ 

-\ Expand[Par[SB[a,b,c,d]**S[i]]**SB[j,k,l]]-

Eta[i,j]\ Par[SB[a,b,c,d]**S[k,l]]+

Eta[i,k]\ Par[SB[a,b,c,d]**S[j,l]]-

Eta[i,l]\ Par[SB        [a,b,c,d]**S[j,k]]

S/:\ S[a\_,b\_,c\_,d\_,e\_]**SB[i\_,j\_,k\_,l\_,m\_]\ :=\ 

Expand[S[a]**Par[S[b,c,d,e]**SB[i,j,k,l,m]]]-

Eta[a,b]\ Par[S[c,d,e]**SB[i,j,k,l,m]]\ +

Eta[a,c]\ Par[S[b,d,e]**SB[i,j,k,l,m]]\ -

Eta[a,d]\ Par[S[b,c,e]**SB[i,j,k,l,m]]\ +

Eta[a,e]\ Par[S[b,c,d]**SB[i,j,k,l,m]]

SB/:\ SB[a\_,b\_,c\_,d\_,e\_]**S[i\_,j\_,k\_,l\_,m\_]\ :=\ 

Expand[S[b,c,d,e]**Par[SB[a]**S[i,j,k,l,m]]]+

Eta[a,b]\ Par[SB[c,d,e]**S[i,j,k,l,m]]\ -

Eta[a,c]\ Par[SB[b,d,e]**S[i,j,k,l,m]]\ +

Eta[a,d]\ Par[SB[b,c,e]**S[i,j,k,l,m]]\ -

Eta[a,e]\ Par[SB[b,c,d]**S[i,j,k,l,m]]

Unprotect[Plus]

x\_?NumberQ\ Epsilon[a\_\_,b\_\_]\ S\_[b\_\_]\ +\ 

y\_?NumberQ\ Epsilon[a\_\_,c\_\_]\ S\_[c\_\_]\ :=\ 

(x+y)\ Epsilon[a,b]**S[b]

x\_?NumberQ\ Epsilon[l\_\_\_,m\_,n\_\_,o\_\_]\ S\_[m\_,o\_\_]\ +\ 

y\_?NumberQ\ Epsilon[l\_\_\_,m\_,n\_\_,p\_\_]\ S\_[m\_,p\_\_]\ :=\ 

(x+y)\ Epsilon[l,m,n,o]**S[m,o]

x\_?NumberQ\ Epsilon[a\_\_,b\_\_]\ S\_[i\_,b\_\_]\ +\ 

y\_?NumberQ\ Epsilon[a\_\_,c\_\_]\ S\_[i\_,c\_\_]\ :=\ 

(x+y)\ Epsilon[a,b]\ S[i,b]

x\_?NumberQ\ Epsilon[l\_\_\_,m\_,n\_\_,o\_\_]\ S\_[j\_,m\_,o\_\_]\ +\ 

y\_?NumberQ\ Epsilon[l\_\_\_,m\_,n\_\_,p\_\_]\ S\_[j\_,m\_,p\_\_]\ :=\ 

(x+y)\ Epsilon[l,m,n,o]\ S[j,m,o]

x\_?NumberQ\ Epsilon[l\_\_\_,m\_,n\_\_,o\_\_]\ S\_[m\_,j\_,o\_\_]\ +\ 

y\_?NumberQ\ Epsilon[l\_\_\_,m\_,n\_\_,p\_\_]\ S\_[m\_,j\_,p\_\_]\ :=\ 

(x+y)\ Epsilon[l,m,n,o]\ S[m,j,o]

Protect[Plus]

(*\ EndPackage[]\ *)

\section*{Acknowledgements}
The authors thanks Eric Sachs, John Veson, and Rachel Williams for
their efforts in the University of Iowa's Summer Science Training
Program. 
%\endmathin

\refs

\end{document}